\documentclass[aps,prb,superscriptaddress,amsmath,amssymb,twocolumn,amsfonts,longbibliography]{revtex4-2}
\usepackage{bm,xcolor,MnSymbol}
\usepackage{graphicx}
\graphicspath{{Pictures/}}
\usepackage{braket,natbib}  
\usepackage[colorlinks=true,linkcolor=magenta,urlcolor=purple,citecolor=magenta,anchorcolor=blue]{hyperref}
\usepackage[normalem]{ulem}
\usepackage{tikz}
\usetikzlibrary{arrows.meta}
\usepackage{orcidlink}

\newcommand{\bs}{\ensuremath{\mathbf{S}}}
\newcommand{\vk}{\ensuremath{\mathbf{k}}}
\newcommand{\vq}{\ensuremath{\mathbf{q}}}
\newcommand{\vp}{\ensuremath{\mathbf{p}}}

\begin{document}

\title{Interacting Dirac magnons in the van der Waals ferromagnet CrBr$_3$}

\author{Saikat Banerjee\,\orcidlink{0000-0002-3397-0308}}
\email{saikat.banerjee@uni-greifswald.de}
\affiliation{Institute of Physics, University of Greifswald, Felix-Hausdorff-Strasse 6, 17489 Greifswald, Germany}
\affiliation{Center for Materials Theory, Rutgers University, Piscataway, New Jersey, 08854, USA}
\author{Stephan Humeniuk\,\orcidlink{0000-0002-7414-8468}}
\email{stephan.humeniuk@gmail.com}
\affiliation{Center for Materials Theory, Rutgers University, Piscataway, New Jersey, 08854, USA}
\date{\today}

\begin{abstract}
We study the effects of magnon-magnon interactions in the two-dimensional van der Waals ferromagnet CrBr$_3$ focusing on its honeycomb lattice structure. Motivated by earlier theoretical predictions of temperature-induced spectral shifts and van Hove singularities in the magnon dispersion~[S. S. Pershoguba \textit{et al}., Dirac Magnons in Honeycomb Ferromagnets, \href{https://journals.aps.org/prx/abstract/10.1103/PhysRevX.8.011010}{Phys. Rev. X {\textbf{8}}, 011010 (2018)}], we go beyond the commonly used thermal magnon approximation by applying second-order perturbation theory in a fully numerical framework. Our analysis uncovers significant deviations from previous analysis: in particular, the predicted singularities are absent, consistent with recent inelastic neutron scattering measurements~[S. E. Nikitin \textit{et al}., Thermal Evolution
of Dirac Magnons in the Honeycomb Ferromagnet CrBr$_3$, \href{https://journals.aps.org/prl/abstract/10.1103/PhysRevLett.129.127201}{Phys. Rev. Lett. {\textbf{129}}, 127201 (2022)}]. Moreover, we find that the temperature dependence of the renormalized magnon spectrum exhibits a distinct $T^3$ behavior for the optical magnon branch, while retaining $T^2$ behavior for the acoustic or down magnon band. This feature sheds new light on the collective dynamics of Dirac magnons and their interactions. We further compare the honeycomb case with a triangular Bravais lattice, relevant for ferromagnetic monolayer MnBi$_2$Te$_4$, and show that both systems lack singular features while displaying quite distinct thermal trends.
\end{abstract}

\maketitle

\section{Introduction \label{sec:sec_I}}

The discovery of graphene~\cite{Geim2007} and topological insulators~\cite{Moore2010} has led to the emergence of a broad class of systems known as Dirac materials, where low-energy fermionic excitations behave like massless relativistic particles with linearly dispersing energy-momentum relations. These so-called Dirac cones are typically protected by symmetries such as time-reversal or inversion~\cite{Wehling2014}, and while the cones are robust to many-body perturbations, interactions among quasiparticles can lead to subtle renormalizations of the spectrum~\cite{Elias2011,RevModPhys.84.1067}. This naturally raises the question of whether similar relativistic-like features can arise in bosonic systems. In recent years, various platforms have indeed been identified as bosonic Dirac materials, including phonons in hexagonal lattices~\cite{Jin2018}, excitations in granular superconductors~\cite{PhysRevB.93.134502}, magnons in honeycomb ferromagnets~\cite{PhysRevB.94.075401,PhysRevB.109.104417}, and even plasmons in topological insulators~\cite{DiPietro2013}. As with their fermionic counterparts, bosonic Dirac dispersions can also be symmetry-protected, yet remain sensitive to the effects of interactions~\cite{PhysRevResearch.2.033035}. A recent theoretical study highlighted how such interactions can qualitatively reshape the spectrum, offering a minimalistic classification of renormalization effects in both the Dirac fermion and boson systems~\cite{Banerjee_2020}.

Among the various bosonic Dirac systems, Dirac magnons have attracted growing attention. The chromium trihalides, CrX$_3$ (X = F, Cl, Br, I), provide a natural platform for realizing such excitations, with CrBr$_3$ and CrCl$_3$ hosting linearly dispersing Dirac magnons~\cite{Burch2018,Wang_2011,Huang2017,Soriano2020}, and CrI$_3$ showing evidence for topologically gapped magnon bands~\cite{Jiang2018,PhysRevX.8.041028,PhysRevX.11.031047,Kim2024}. These are magnetic van der Waals materials where spins reside on a two-dimensional honeycomb lattice, making them ideal for exploring magnon band topology. In an earlier theoretical work~\cite{PhysRevX.8.011010}, one of us studied Dirac magnons in ferromagnetic CrBr$_3$ and demonstrated that magnon-magnon interactions lead to temperature and momentum-dependent renormalization of the Dirac dispersion, captured within a generalized Dyson spin-wave framework~\cite{PhysRev.102.1217,PhysRev.102.1230}. Notably, we found a characteristic $T^2$ temperature dependence in the renormalized magnon energies, along with prominent van Hove-like structures in the renormalized spectrum, especially near the M points in the Brillouin zone (BZ). These findings were motivated by longstanding anomalies observed in earlier inelastic neutron scattering (INS) experiments on CrBr$_3$~\cite{PhysRevB.4.2280,PhysRevB.3.157}.
    
In this previous work~\cite{PhysRevX.8.011010}, we adopted a key approximation concerning two-particle scattering. One magnon was assumed to be thermally excited near the bottom of the magnon spectrum, while the other, which can be anywhere in the BZ, was induced by the incoming neutron. This led to a subtle magnonic band renormalization as mentioned earlier. However, recent INS measurements by Nikitin et al.~\cite{PhysRevLett.129.127201} on CrBr$_3$ have revealed a discrepancy with these earlier predictions. While the study~\cite{PhysRevLett.129.127201} confirmed $T^2$- band renormalization, the predicted momentum-dependent van Hove features remained unobserved. Dip structures in the magnon spectrum around the BZ boundaries have also been predicted for ferromagnetic Bravais lattices, including the monolayer ferromagnet MnBi$_2$Te$_4$. However, these features have not been observed in other INS experiments~\cite{Bayrakci2006, PhysRevLett.111.017204}, necessitating a more sophisticated theoretical investigation. 

The investigation of magnon renormalization in both three-dimensional and two-dimensional ferromagnets has a long history. While Dysonian spin-wave theory~\cite{PhysRev.102.1217,PhysRev.102.1230} discusses the interaction and associated perturbative corrections within a long-wavelength limit, it is crucial to retain the complete momentum-resolved interaction vertices, especially for two-dimensional magnets. This necessity arises from the enhanced spin fluctuations that occur in two dimensions. Earlier theoretical work~\cite{PhysRevB.60.1082} has demonstrated that going beyond first-order perturbation theory in a self-consistent manner is essential to accurately capture the temperature dependence of magnetization in layered compounds. Numerous studies have employed various versions of self-consistent spin wave theory~\cite{Irkhin_1992}, equation-of-motion-based spin Green's functions~\cite{PhysRevB.54.1019}, and other techniques to effectively address magnon renormalization. A comprehensive and detailed reference on spin-wave renormalization can be found in works such as Refs.~\cite{Lee2013,RevModPhys.95.035004}.

Motivated by these experimental observations, here we revisit the interacting Dirac magnon in CrBr$_3$, going beyond the \textit{thermal magnon approximation} \--- a key approximation adopted earlier. We further incorporate a small on-site single-ion anisotropy to reconcile the observed long-range magnetic order at finite temperature with the Mermin-Wagner theorem~\cite{PhysRevLett.17.1133}. To capture the renormalization in an unbiased manner, we perform extensive numerical computations to evaluate self-consistent self-energy corrections upto second order in perturbation theory. The crucial difference from our previous work~\cite{PhysRevX.8.011010} lies in the fact that we retain the full momentum-resolved interaction vertices, and perform the momentum summations over the honeycomb BZ without adhering to any long-wave length approximations. Moreover, the first-order (Hartree) corrections are incorporated into the second-order self-energy in a self-consistent manner which leads to qualitative changes in the magnon renormalization profile. Finally, we obtain the emergence of negative renormalization in the second-order perturbation theory which remains valid as long as it does not exceed the single ion anisotropy for low temperatures. However, the negative magnitude increases with increasing temperature and therefore puts a strong constraint for the applicability of perturbation theory at high temperatures.

We find that the renormalized magnon spectrum lacks van Hove singularities near the BZ corners, although exhibiting a strong, non-monotonic momentum dependence. A direct comparison with recent INS measurement~\cite{PhysRevLett.129.127201} supports our findings while contrasting sharply with previous theoretical predictions of van Hove like features from multiple previous works~\cite{PhysRevX.8.011010,YiqunLiu_PRB2023_MnBi2Te4,PhysRevB.4.2280}. Our work emphasizes that retaining the full momentum-dependent vertex and self-energy structure is a crucial aspect leading to the absence of van Hove peaks in the renormalized magnon spectrum.

Analysis of the temperature dependence of the magnon energy renormalization reveals a deviation from the previously predicted $T^2$- behavior.  While our numerical results remain consistent with a $T^2$- dependence for the acoustic magnon branch, the optical branch exhibits a distinct $T^3$- dependence. Notably, a recent theoretical work~\cite{PhysRevB.111.094430} also reported a $T^3$- dependence across the entire renormalized magnon band structure for CrI$_3$. In contrast, we find quadratic temperature dependence for the acoustic branch and cubic dependence for the optical branch of the honeycomb lattice.  We note that small fluctuations in the temperature exponent $\alpha$ for both up ($\alpha = 3 \pm \delta)$ and down $(\alpha = 2 \pm \delta$) bands persist across the entire BZ. We interpret these fluctuations as arising from thermal effects~\cite{Kosevich1986}. Further controlled INS experiments are needed to confirm this modified temperature dependence and elucidate the underlying physics. The absence of van Hove features in magnon renormalization near the BZ boundary, and a distinct $T^3$ dependence of the second-order self-energy correction in the optical magnon branch are the two primary results of our work in honeycomb ferromagnets.

We also performed a similar analysis for the triangular lattice, relevant to monolayer MnBi$_2$Te$_4$, and confirmed that a second-order perturbation theory \--- without the thermal magnon approximation \--- does not produce any van Hove singularities or dip structures, consistent with prior experimental observations~\cite{Bayrakci2006, PhysRevLett.111.017204}. On the other hand, the second-order perturbation correction hosts a $T^2 \log T$ temperature dependence, quite contrary to the previous non-Bravais case. We believe this is a crucial difference in the renormalized  magnon spectra between Bravais and non-Bravais lattices.

The remainder of this paper is organized as follows. Sec.~\ref{sec:sec_II} revisits the non-interacting magnons within a ferromagnetic Heisenberg model and presents the interaction terms that arise beyond the linear spin-wave approximation. We then analyze the Hartree and second-order perturbation theory in Sec.~\ref{sec.sec_II.I} and Sec.~\ref{sec.sec_II.II}, respectively, assuming large spin ($S = \tfrac{3}{2}$ for CrBr$_3$). Next, in Sec.~\ref{sec:sec_III}, we compare our numerical results with the experimental findings from the INS study reported in Ref.~\cite{PhysRevLett.129.127201}. The analysis for the triangular lattice ferromagnet MnBi$_2$Te$_4$ is provided in Sec.~\ref{sec:sec_IV}. Finally, we conclude in Sec.~\ref{sec:sec_V}.

\section{Model \label{sec:sec_II}}

We start with a ferromagnetic Heisenberg model on a honeycomb lattice with single-ion anisotropy relevant for CrBr$_3$. The corresponding Hamiltonian is written as 
\begin{equation}\label{eq.1}
\mathcal{H} = - J \sum_{\langle ij \rangle}  \bs_i \cdot \bs_j + A \sum_i (S^z_i)^2,
\end{equation}
where $J$ is the strength of the ferromagnetic coupling, and $A$ is the single-ion anisotropy term. We adopt the parameters from Ref.~\cite{PhysRevLett.129.127201} and choose $J = 1.494$ meV and $A = -0.028$ meV. The tiny magnitude of the single-ion anisotropy term is adopted from the previous ferromagnetic resonance experiment~\cite{PhysRevB.56.719}. 

A few comments are necessary here regarding the validity of the specific model adopted here. In principle, the CrX$_3$ for all three halides (X = Cl, Br, I) are characterized by $d$-orbital excitations as $^4A_2 (t^3_{2g}-e^0_{g})$ with the atomic ground state lying $\sim 1.5 - 1.7$ eV below the excited state $^4T_2(t^2_{2g}-e^1_{g})$ \--- leading to an effective spin angular momentum as $S = \tfrac{3}{2}$~\cite{Yadav2024}. Previous theoretical works have elucidated the role of $t_{2g}-e_g$ interactions~\cite{PhysRevMaterials.3.031001}, the role of ligand orbitals~\cite{PhysRevB.99.104432,Soriano2021}, and stacking sequence~\cite{Wang2021} to stabilize the ferromagnetic order. The larger the ligand atom in size, the more relevant are the other interactions in the spin exchange Hamiltonian. For example, it was shown in Ref.~\cite{PhysRevResearch.3.013216} that the large spin-orbit coupling in I atom leads to significant Kitaev interactions in the spin-exchange model mediated by ligand-transition metal hopping. On the other hand, $t_{2g}-e_g$ mixing~\cite{PhysRevB.97.085150} favors anisotropic exchange coupling leading to the well-known $J-K-\Gamma$ model. Furthermore, any dipolar interactions, if present, can lead to in-plane single ion anisotropy which can modify the above Hamiltonian in subtle manner~\cite{Yadav2024}. In the context of CrBr$_3$, however, the other exchange couplings are much smaller compared to CrI$_3$ as revealed by recent \textit{ab initio} calculations~\cite{PhysRevB.103.125418}. Hence, the above model in Eq.~\eqref{eq.1} serves as a good starting point for our renormalization analysis. At this point, we note that a contemporary INS experiment on CrBr$_3$~\cite{PhysRevB.104.L020402} reported the presence of a large gap a the Dirac point due to the presence of a strong Dzyaloshinskii-Moriya (DM) interaction. Whether such DM interaction is present in CrBr$_3$ or not is a controversial question. Without aiming to resolve this controversy, we do not consider such DM terms and continue with the model in Eq.~\eqref{eq.1}.

\begin{figure}[t!]
\centering \includegraphics[width=1.0\linewidth]{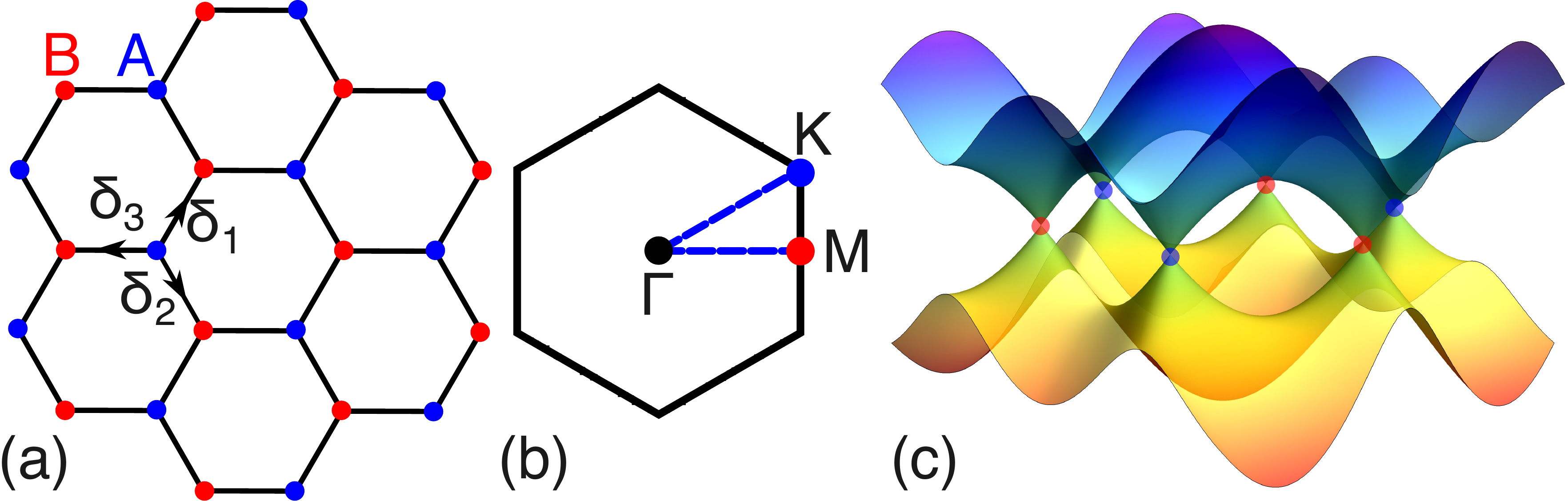}
\caption{(a) The honeycomb lattice for the ferromagnetic Heisenberg model. Red (blue) symbols signify the inequivalent lattice sites A (B), with $\bm{\delta}_i$’s being its three nearest-neighbor vectors. (b) The corresponding Brillouin zone (BZ) with the high-symmetry points, labeled by the colored dots. (c) Non-interacting Dirac magnons with linearly dispersing spectrum at the K, and K' points in the BZ. The lattice coordinates of the high-symmetry points are given as K = ($\tfrac{2\pi}{3a}, \tfrac{2\pi}{3\sqrt{3}a}$), and M = ($\tfrac{2\pi}{3a},0$), where $a$ is the lattice constant.}\label{fig:Fig1}
\end{figure}

Much below the Curie temperature ($T_c \sim 37$ K)~\cite{Tang2020,Fumega2020,Kozlenko2021,Zhang2024}, CrBr$_3$ is well described by linear spin-wave theory using the Holstein-Primakoff transformation for the spin operators. The latter is defined as 
\begin{equation}\label{eq.2}
S^{+}_i \approx \sqrt{2S} \left( 1 - \frac{n_{\eta_i} }{4S}\right) \eta_i, \;
S^{-}_i = (S^{+}_i)^\dag, \;
S^{z}_i = S - n_{\eta_i},
\end{equation}
where $S^{\pm}_i = S^{x}_i \pm i S^{y}_i$ are the ladder operators, $n_{\eta_i} = \eta^\dag_i \eta_i$ is the number operator, $S = \frac{3}{2}$, and $\eta_i = (a_i, b_i)$ are the corresponding magnonic operators defined on the two sub-lattices on a honeycomb lattice. Expanding the spin operators up to leading order in the bosonic expansion and rewriting in the momentum space, we obtain the non-interacting Dirac Hamiltonian as
\begin{equation}\label{eq.3}
\mathcal{H}_0 = JS 
\sum_{\vk} 
\begin{pmatrix}
a^\dag_\vk	&	b^\dag_\vk 
\end{pmatrix}
\begin{pmatrix}
3 - \frac{2A}{J}	&	-\gamma_\vk \\
-\gamma^\ast_\vk	&	3 - \frac{2A}{J}
\end{pmatrix}
\begin{pmatrix}
a_\vk \\
b_\vk
\end{pmatrix},
\end{equation}\\
where $\gamma_\vk = \sum_i e^{i \vk \cdot \bm{\delta}_i}$ with $\bm{\delta}_i = \{\bm{\delta}_1, \bm{\delta}_2, \bm{\delta}_3\}$ being the three nearest-neighbor vectors in the honeycomb geometry as shown in Fig.~\ref{fig:Fig1}(a). The neighboring vectors are chosen as $\bm{\delta}_1 = \tfrac{a}{2} (1, \sqrt{3})$, $\bm{\delta}_2 = \tfrac{a}{2} (1, -\sqrt{3})$, and $\bm{\delta}_3 = a(-1,0)$. The above Hamiltonian can be diagonalized to obtain the dispersion relations for the optical ``up" and the acoustic ``down" magnons as $\epsilon^{u,d}_\vk = JS (3 - \tfrac{2A}{J} \pm |\gamma_\vk|)$. Here, up and down branch magnonic wave functions are defined as $\Psi^{u,d}_\vk = (e^{i\phi_\vk/2} \mp e^{-i\phi_\vk/2})/\sqrt{2}$, and the momentum-dependent phase is defined as $\phi_\vk = {\rm{arg}} \, \gamma_\vk$. The corresponding spectrum for \textit{up} and \textit{down} magnons is shown in Fig.~\ref{fig:Fig1}(c).

This description is valid at very low temperatures where the number of thermally excited magnons is small. However, at higher temperatures ($T \lesssim T_c$), a non-interacting description becomes inadequate, and we must account for magnon-magnon interactions. These interactions naturally arise from the Holstein-Primakoff expansion in next-to-leading order. Following Ref.~\cite{PhysRevX.8.011010}, we express the interacting magnon Hamiltonian in the diagonal basis as $\mathcal{H} = \mathcal{H}_0 + \mathcal{H}_1$, with the interacting part given by
\begin{widetext}
\begin{align}
\nonumber
\mathcal{H}_1
=
\frac{J}{4N} 
\sum_{\vk,\vq,\vp} 
&
\bigg[
\Big[
(|\gamma_\vq|-|\gamma_\vk|) c_\zeta 
+
|\gamma_{\vp-\vq}| c_\theta
-
|\gamma_{\vp-\vk}| c_\kappa
\Big] 
u^\dag_\vp d^\dag_{\vk+\vq-\vp} d_\vq u_\vk
+
i 
\Big[
(|\gamma_\vk|-|\gamma_\vq|) s_\zeta
+ 
|\gamma_{\vp-\vq}| s_\theta 
+
|\gamma_{\vp-\vk}| s_\kappa
\Big] d^\dag_\vp d^\dag_{\vk+\vq-\vp} d_\vq u_\vk \\
\nonumber
&
+
\Big[ 
(|\gamma_\vk| + |\gamma_\vq|) c_\zeta
-
|\gamma_{\vp-\vq}| c_\theta
-
|\gamma_{\vp-\vk}| c_\kappa
\Big] 
d^\dag_\vp d^\dag_{\vk+\vq-\vp} d_\vq d_\vk 
-
\Big[ 
(|\gamma_\vk| + |\gamma_\vq|) c_\zeta
-
|\gamma_{\vp-\vq}| c_\theta
-
|\gamma_{\vp-\vk}| c_\kappa
\Big] 
u^\dag_\vp u^\dag_{\vk+\vq-\vp} u_\vq u_\vk \\ 
\label{eq.4}
&
+
\frac{1}{2} 
\Big[
(|\gamma_\vk| + |\gamma_\vq|) e^{i\zeta}
- 
2 i|\gamma_{\vp-\vq}| s_\theta
-
2i|\gamma_{\vp-\vk}| s_\kappa
\Big] d^\dag_\vp u^\dag_{\vk+\vq-\vp} u_\vq u_\vk
\bigg] + \rm{h.c.},
\end{align}
\end{widetext}
where $c_x = \cos x$, $s_x = \sin x$, $N$ is the number of sites, and the three angles in the scattering amplitudes are defined as 
\begin{widetext}
\begin{equation}\label{eq.5}
\zeta	 = \frac{\phi_\vp + \phi_{\vk+\vq-\vp} - \phi_\vq - \phi_\vk}{2}, \quad
\theta	 = \frac{2\phi_{\vp-\vq} - \phi_\vp + \phi_{\vk+\vq-\vp} + \phi_\vq - \phi_\vk}{2}, \quad
\kappa	 = \frac{2\phi_{\vk-\vp} + \phi_\vp - \phi_{\vk+\vq-\vp} + \phi_\vq - \phi_\vk}{2}.
\end{equation}
\end{widetext}
We ignored the interaction contribution from the single-ion anisotropy term, assuming $J \gg A$. Furthermore, the other longer-range hopping terms, such as $J_2 = 0.077$ meV and $J_3 = -0.068$ meV, are also ignored in consideration of $J \gg J_2, J_3$~\cite{PhysRevLett.129.127201}. Although retaining all hopping terms is necessary for a complete quantitative analysis, we present a qualitative comparison. The hierarchy $J \gg J_2, J_3, A$ suggests our results will nonetheless exhibit good quantitative agreement.

Now, we analyze the effect of these interactions on the bare magnon bands using perturbation theory. Consequently, we focus on the two lowest-order contributions in perturbation using the Hartee and the Sunset diagrams as shown in Fig.~\ref{fig:Fig2}(a,b). Let us first analyze the Hartree diagram in the following section. \\

\subsection{Hartree contribution \label{sec.sec_II.I}}

\begin{figure}[b!]
\centering \includegraphics[width=0.8\linewidth]{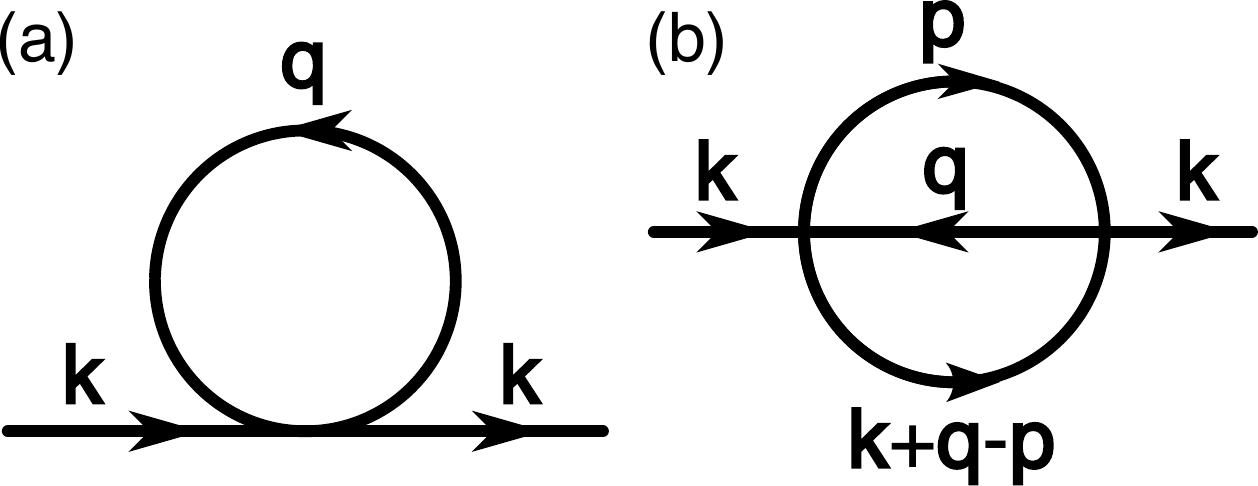}
\caption{A schematic for the (a) Hartree, \textit{i.e.}, in the first order perturbation, and (b) Sunset, \textit{i.e.}, in the second order perturbation diagram with magnon-magnon scattering interactions given in Eq.~\eqref{eq.4}.}\label{fig:Fig2}
\end{figure}

Considering spin $S = 3/2$, we set up the perturbation theory in up and down magnon branches. The first-order Hartree correction is computed from the Feynman diagram in Fig.~\ref{fig:Fig2}(a). Following Ref.~\cite{PhysRevX.8.011010}, we obtain the corresponding self-energy contribution in the sub-lattice basis as [see Appendix~\ref{sec:sec_app.I}]
\begin{equation}\label{eq.6}
\Sigma_{\rm{H}}(T) = \frac{J}{2N}
\sum_{\vk} 
\begin{bmatrix}
a^\dag_\vk	&	b^\dag_\vk 
\end{bmatrix}
\begin{bmatrix}
h(T)	&	g_\vk(T) \\
g^\ast_\vk(T)	&	h(T)
\end{bmatrix}
\begin{bmatrix}
a_\vk \\
b_\vk
\end{bmatrix}, 
\end{equation}
where the temperature-dependent functions $h(T)$, and $g_\vk(T)$ are defined as 
\begin{widetext}
\begin{equation}\label{eq.7}
h(T) = - \sum_\vq 
\big[
(|\gamma_\vq| + |\gamma_0|) f(\epsilon^u_\vq)
- 
(|\gamma_\vq| - |\gamma_0|) f(\epsilon^d_\vq)
\big], 
\quad
g_\vk(T) = 
\sum_\vq 
\big[
(\gamma_\vk + e^{i \phi_\vq}\gamma_{\vk-\vq}) f(\epsilon^u_\vq) 
+
(\gamma_\vk - e^{i \phi_\vq}\gamma_{\vk-\vq}) f(\epsilon^d_\vq)
\big].
\end{equation}
\end{widetext}
Here, $f(x) = 1/(e^{\beta x}-1)$ is the Bose-Einstein distribution function with $\beta = 1/k_{\rm{B}}T$. The summation over momenta in Eq.~\eqref{eq.7} is defined in the honeycomb lattice BZ. Note the additional terms related to the optical magnon band-structure in $h(T)$ and $g_\vk(T)$ in comparison to Ref.~\cite{PhysRevX.8.011010}. We retain the full band structure for the up and the down magnons in Eq.~\eqref{eq.7} as shown in Fig.~\ref{fig:Fig1}(c), and perform the momentum integration numerically using the conventional quadrature rule~\cite{Bulirsch2002}.

\begin{figure}[b!]
\centering \includegraphics[width=1.0\linewidth]{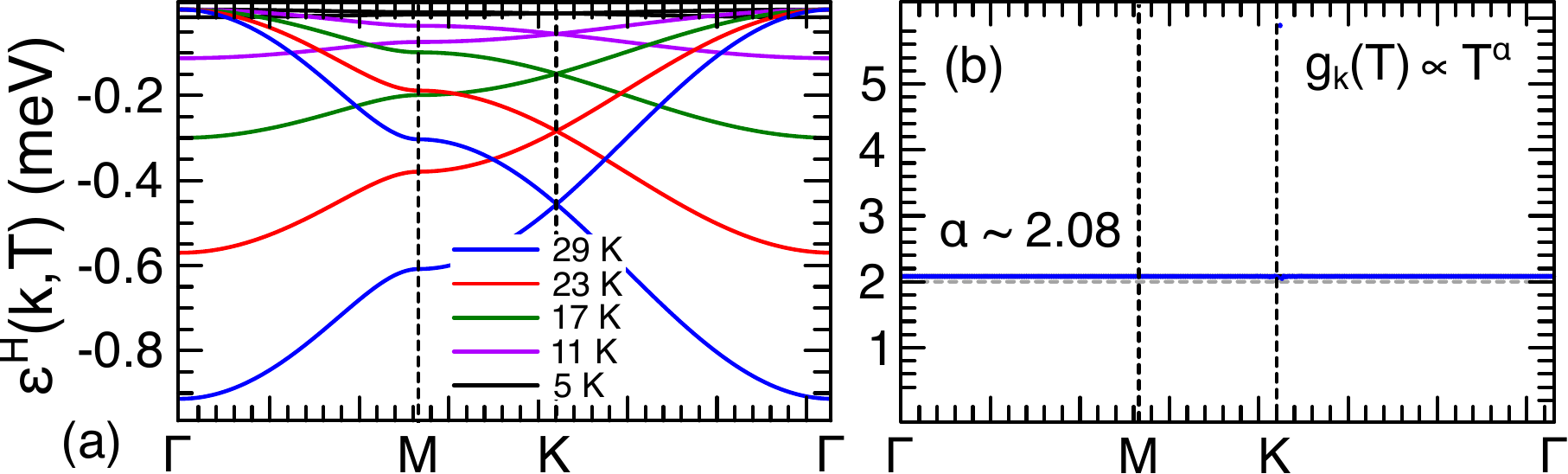}
\caption{(a) Temperature-dependent Hartree spectrum, calculated by diagonalizing the Hartree term in Eq.~\eqref{eq.6}. (b) Best-fit exponent $\alpha$ obtained by fitting the temperature dependence of $g_\vk(T) \sim T^\alpha$ to numerical data obtained at fourteen equidistant temperatures between 5K and 31K.}\label{fig:Fig3}
\end{figure}

To capture the temperature dependence of both functions $h(T)$ and $g_\vk(T)$, we perform numerical integration as outlined in Eq.~\eqref{eq.7} within the honeycomb BZ. We first focus on the Hartree spectrum by diagonalizing Eq.~\eqref{eq.6} at various temperatures, as shown in Fig.~\ref{fig:Fig3}(a). The resulting spectrum qualitatively resembles the bare magnon spectrum in Fig.~\ref{fig:Fig1}(c), but exhibits significant bandwidth renormalization. To determine the temperature dependence of this renormalization, we evaluate $h(T)$ and $g_\vk(T)$ at high-symmetry points in the BZ [Fig.~\ref{fig:Fig3}(b)] and perform a polynomial fit of the form $h(T) \sim T^\alpha$ and $g_\vk(T) \sim T^\alpha$. We obtain the numerical values for both $h(T)$, and $g_\vk(T)$ at fourteen different temperatures ranging from 5K to 31K. The best fit yields $\alpha = 2.08$, confirming the previously predicted $T^2$ temperature dependence of the Hartree correction. The small deviation $(\approx 0.08)$ from the expected quadratic dependence is probably due to the choice of fit range in temperature or a plausible log-correction as predicted in Ref.~\cite{Kosevich1986}.

\subsection{Sunset contribution \label{sec.sec_II.II}}

In this section, we focus on the second-order perturbation correction governed by the five distinct scattering processes as written in Eq.~\eqref{eq.4}. The associated Feynman diagram is illustrated in Fig.~\ref{fig:Fig2}(b) for the third process. For other processes, the Green's function lines must be changed depending upon ``up" or ``down" magnon types. Following Ref.~\cite{PhysRevX.8.011010}, we obtain the self-energy correction with the interaction defined in Eq.~\eqref{eq.4} in the diagonal basis as [note that we defined the integral $\int_{\vq \vp}$ as $\int_{\vq\vp} = \int_{\text{\tiny{BZ}}} \tfrac{d \vq d \vp}{V_{\text{\tiny{BZ}}}^2}$, where $V_{\text{\tiny{BZ}}} = \tfrac{8\pi^2}{3\sqrt{3}}$ is the area of the honeycomb BZ as shown in Fig.~\ref{fig:Fig1}(b)]
\begin{equation}\label{eq.9}
\Sigma_{\rm{S}}(\omega,\vk)
=
\frac{J^2}{16} \int_{\vq\vp}
\frac{{|\cal{V}_{\vk;\vq,\vp}|}^2 \mathcal{F}_{\vk;\vq,\vp}}{\omega + \varepsilon_\vq - \varepsilon_\vp - \varepsilon_{\vk+\vq-\vp} + i\delta},
\end{equation}
where $\cal{V}_{\vk;\vq,\vp}$ is the matrix element belonging to one of the five processes, $\delta$ is a small positive number, and $\varepsilon_\vk$ corresponds to either up or down magnon as dictated by each process in Eq.~\eqref{eq.4}. Finally, the temperature dependent factor $\mathcal{F}_{\vk;\vq,\vp}$ is obtained as [see Appendix.~\ref{sec:sec_app.II}]
\begin{align}
\nonumber
{\mathcal F}_{\vk;\vq,\vp} = & 
f(\varepsilon_\vq) + 
f(\varepsilon_\vq) f(\varepsilon_\vp) \, + \,\\
&
\label{eq.10}
 f(\varepsilon_\vq) f(\varepsilon_{\vk+\vq-\vp})
-
f(\varepsilon_\vp)f(\varepsilon_{\vk+\vq-\vp}).
\end{align}
Now instead of performing the thermal magnon approximation as done in Ref.~\cite{PhysRevX.8.011010}, we retain the full expression for $\mathcal{F}_{\vk;\vq,\vp}$, and $\mathcal{V}_{\vk;\vq,\vp}$, and evaluate the contribution by performing four-dimensional momentum integration in the honeycomb BZ as shown in Fig.~\ref{fig:Fig1}(b). The integrations are performed numerically using Simpson's 1/3 rule~\cite{Atkinson1989}. In the subsequent five paragraphs, we discuss each of the five processes, respectively. To obtain the self-energy corrections, we utilize the on-shell condition \textit{i.e.}, replace $\omega \rightarrow \varepsilon_\vk$ for each of the corresponding five processes. \\

\begin{figure}[t!]
\centering \includegraphics[width=1.0\linewidth]{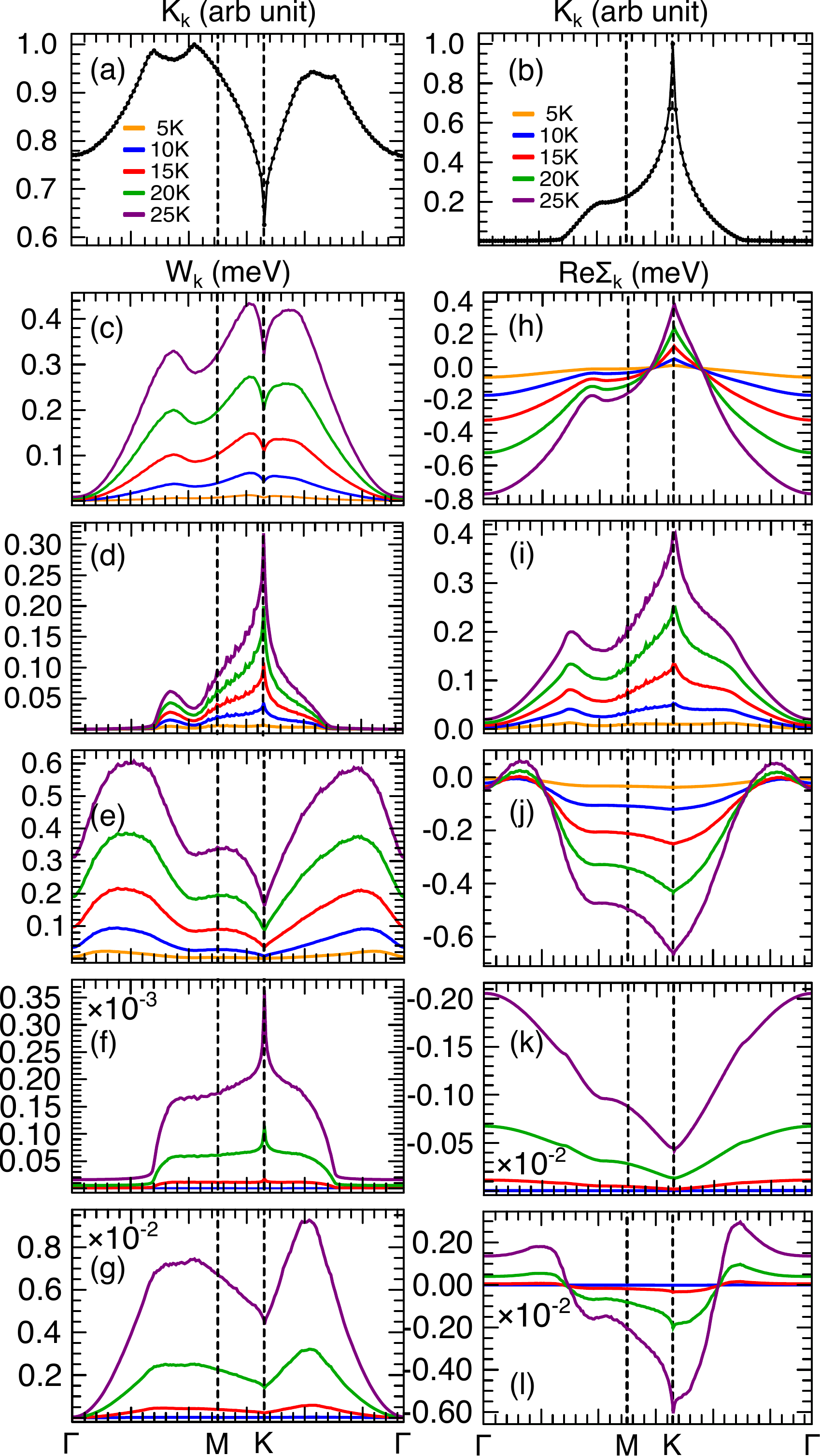}
\caption{The kinematic factors, depicted as solid black lines [in panels (a) and (b)], along with the corresponding decay rates, represented as dashed blue lines, are outlined for each process discussed in Sec.~\ref{sec.sec_II.II}. The processes, labeled from first to fifth, are illustrated in panels (c) through (g). Panels (h) through (l) present the real part of the self-energy for the processes ranked first to fifth, respectively. The results are obtained at various temperatures as mentioned in panel (a,b). The magnitudes are in meV with the parameters chosen as $J = 1.494$ meV and $A = 0.028$ meV. The first and fifth processes have similar kinematic factor as in (a), while the remaining three processes have similar kinematic factors as in (b).}\label{fig:Fig4}
\end{figure}

\paragraph{\textbf{First process} \underline{$d + d \rightarrow d + d$}\textbf{:}} We begin by analyzing the most dominant magnon decay process, which involves only the down-band magnons. This channel constitutes the leading contribution due to thermal population factors. The matrix element for this process, derived in Eq.~\eqref{eq.4}, takes the form
\begin{equation}\label{eq.11}
{\cal{V}}^{(1)}_{\vk;\vq,\vp} = 
(|\gamma_\vk| + |\gamma_\vq|) c_\zeta -
|\gamma_{\vp-\vq}| c_\theta -
|\gamma_{\vp-\vk}| c_\kappa,
\end{equation}
while the associated kinematic factor is given by
\begin{equation}\label{eq.12}
K^{(1)}_\vk = \int_{\vq\vp} \delta(\epsilon^d_\vk + \epsilon^d_\vq - \epsilon^d_\vp - \epsilon^d_{\vk+\vq-\vp}).
\end{equation}
The decay rate for this low-energy process is then obtained from Eq.~\eqref{eq.9} as
\begin{equation}\label{eq.13}
W^{(1)}_\vk = \int_{\vq\vp} \frac{J^2|\mathcal{V}^{(1)}_{\vk;\vq,\vp}|^2 {\mathcal F}_{\vk;\vq,\vp}}{16} \delta(\epsilon^d_\vk + \epsilon^d_\vq - \epsilon^d_\vp - \epsilon^d_{\vk+\vq-\vp}).
\end{equation}
The resulting decay profile is shown in Fig.~\ref{fig:Fig4}(c), while the corresponding kinematic factor $K_\vk^{(1)}$ is displayed in Fig.~\ref{fig:Fig4}(a). It is evident that the matrix elements, along with the temperature-dependent Bose distribution factors, induce notable qualitative differences in the decay rate \--- particularly near high-symmetry points in the BZ \--- compared to our earlier theoretical results in Ref.~\cite{PhysRevX.8.011010}. The real part of the associated self-energy correction is shown in Fig.~\ref{fig:Fig4}(h).

\begin{figure}[t!]
\centering \includegraphics[width=1.0\linewidth]{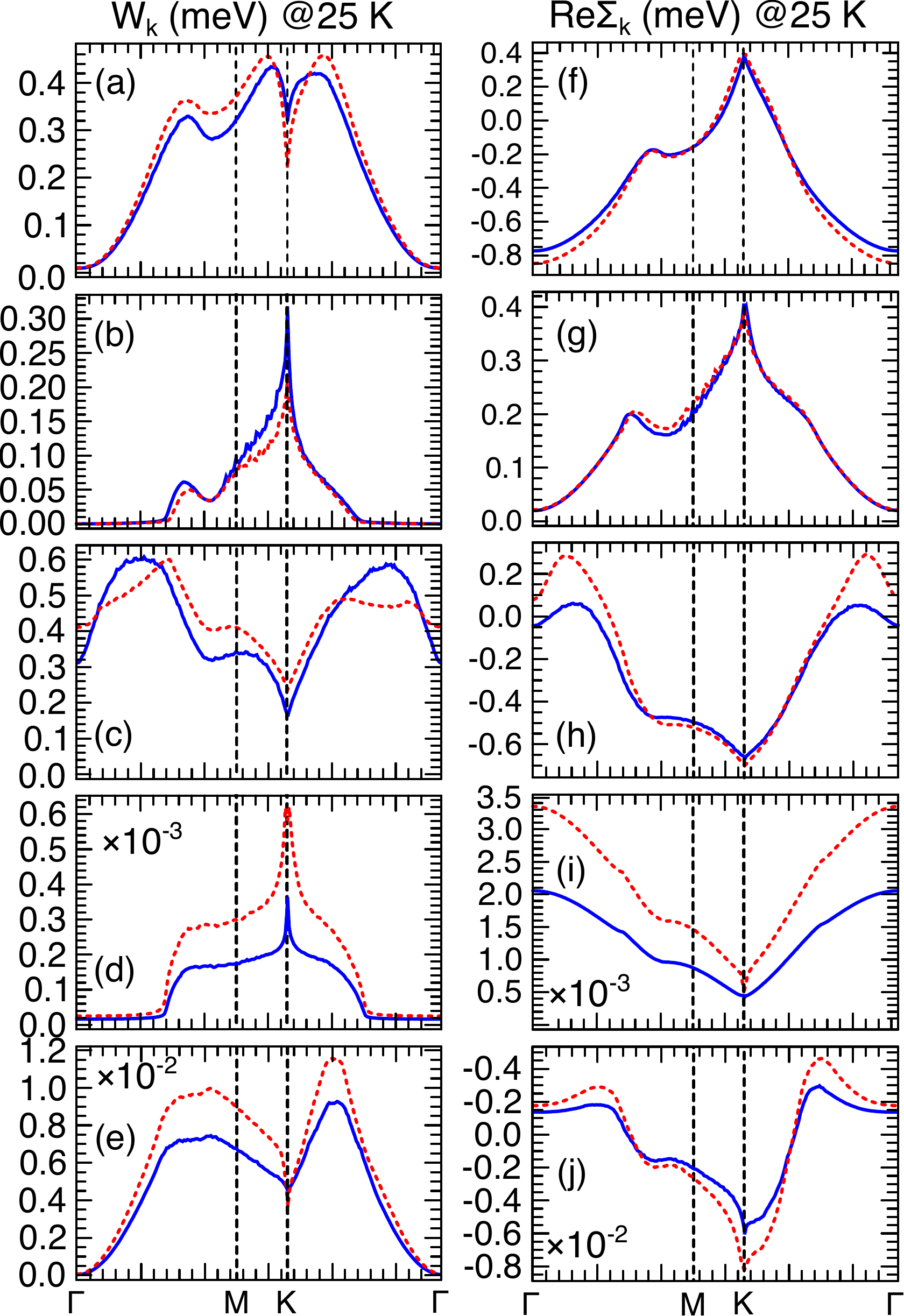}
\caption{Comparison between bare and self-consistent second-order perturbation results at higher temperature (25 K). The solid lines represent the bare second-order corrections, while the dashed lines correspond to the self-consistent results, where the Hartree correction is incorporated into the magnon energies. Panels (a–e) display the decay rates for processes 1 through 5, and panels (f–j) show the corresponding real parts of the self-energy. Notably, the self-consistent treatment leads to significant qualitative modifications in both decay rates and real parts, including changes in the relative magnitudes across the BZ.}\label{fig:Fig5}
\end{figure}

For completeness, we also carry out a self-consistent perturbative analysis, incorporating the Hartree energy shift into the second-order self-energy correction. This self-consistent approach significantly modifies both the imaginary and real parts of the self-energy at elevated temperatures, as shown in panels (a) and (f) of Fig.~\ref{fig:Fig5}, respectively. At lower temperatures, however, the deviation from the non-self-consistent result remains small. Near the BZ $\Gamma$ point, the real part of the second-order self-energy becomes negative, and its magnitude increases with temperature. This is permissible if the total renormalized energy, $\varepsilon^{{\rm R}}_\Gamma = 2AS + \varepsilon^{(\rm H)}_{\Gamma} + {\rm Re}\,\Sigma^{(1)}_{\rm S} (\Gamma)$, remains positive definite. Consequently, this result limits the range of validity of the perturbation theory, which is applicable at low temperatures but may not be relevant at higher temperatures. \\

\paragraph{\textbf{Second process} \underline{$u + d \rightarrow d + d$}\textbf{:}} In this process, there is a mismatch between the flavors of the incoming and the outgoing magnons, and this can be thought of as an effective scattering channel where an acoustic magnon is converted into an optical magnon. The matrix element is given by
\begin{equation}\label{eq.14}
{\cal{V}}^{(2)}_{\vk;\vq,\vp}
=
i 
(|\gamma_\vk|-|\gamma_\vq|) s_\zeta
+ 
i |\gamma_{\vp-\vq}| s_\theta 
+
i |\gamma_{\vp-\vk}| s_\kappa,
\end{equation}
while the kinematic factor for this case is given by following Eq.~\eqref{eq.4} as
\begin{equation}\label{eq.15}
K^{(2)}_\vk = \int_{\vq\vp} \delta(\epsilon^u_\vk + \epsilon^d_\vq - \epsilon^d_\vp - \epsilon^d_{\vk+\vq-\vp}).
\end{equation}
The corresponding result is shown in Fig.~\ref{fig:Fig4}(b). On the other hand, the decay rate defined by 
\begin{equation}\label{eq.16}
W^{(2)}_\vk = \int_{\vq\vp} \frac{J^2|\mathcal{V}^{(2)}_{\vk;\vq,\vp}|^2 {\mathcal F}_{\vk;\vq,\vp}}{16} \delta(\epsilon^u_\vk + \epsilon^d_\vq - \epsilon^d_\vp - \epsilon^d_{\vk+\vq-\vp}),
\end{equation}
is shown in Fig.~\ref{fig:Fig4}(d) at various temperatures. Matrix elements and temperature factors indeed \textit{significantly} alter the qualitative profile of the decay rate. Most importantly, we do not find any significant divergences near the M point, as was previously pointed out in Ref.~\cite{PhysRevX.8.011010}. The corresponding real part of the self-energy contribution to this process is shown in panel (i) of Fig.~\ref{fig:Fig4}. 

As before, we perform a self-consistent analysis of the second-order self-energy by incorporating the Hartree correction into the bare magnon energies. The resulting imaginary and real parts of the self-energy are shown in panels (b) and (g) of Fig.~\ref{fig:Fig5}, respectively, at a higher temperature. This comparison highlights the importance of including self-consistent analysis for quantitative accuracy. \\

\paragraph{\textbf{Third process} \underline{$u + d \rightarrow u + d$}\textbf{:}} In contrast to the previous scattering channel, here the flavor number is preserved between the incoming and the outgoing magnons. The matrix element for this process is given by [see Eq.~\eqref{eq.4}]
\begin{equation}\label{eq.17}
{\cal{V}}^{(3)}_{\vk;\vq,\vp}
=
(|\gamma_\vq|-|\gamma_\vk|) c_\zeta 
+
|\gamma_{\vp-\vq}| c_\theta
-
|\gamma_{\vp-\vk}| c_\kappa,
\end{equation}
where the angles ($\zeta, \theta, \kappa$) are defined as in Eq.~\eqref{eq.5}. To analyze the associated self-energy contribution in this case, we focus on the kinematic factor determined by the denominator in Eq.~\eqref{eq.9} as 
\begin{equation}\label{eq.18}
K^{(3)}_\vk = \int_{\vq\vp} \delta(\epsilon^u_\vk + \epsilon^d_\vq - \epsilon^u_\vp - \epsilon^d_{\vk+\vq-\vp}).
\end{equation}
The corresponding kinematic factor  is illustrated in Fig.~\ref{fig:Fig4}(b). Going beyond thermal magnon approximation already leads to significant qualitative changes in the kinematic factor in stark contrast to our previous predictions~\cite{PhysRevX.8.011010}. We now compare the decay rate with the kinematic factor. The latter is written as 
\begin{equation}\label{eq.19}
W^{(3)}_\vk = \int_{\vq\vp} \frac{J^2|\mathcal{V}^{(3)}_{\vk;\vq,\vp}|^2 {\mathcal F}_{\vk;\vq,\vp}}{16} \delta(\epsilon^u_\vk + \epsilon^d_\vq - \epsilon^u_\vp - \epsilon^d_{\vk+\vq-\vp}).
\end{equation}
The contrast between the kinematic factor and the decay rate is evident in a comparison of Fig.~\ref{fig:Fig4}(b) and (e). The real part of the corresponding self-energy is obtained numerically, and the resulting spectrum is shown in Fig.~\ref{fig:Fig4}(j) at various temperatures. 

The self-consistent treatment in this case again leads to qualitative changes both in the decay rate and real part as illustrated in Fig.~\ref{fig:Fig5}(c) and Fig.~\ref{fig:Fig5}(h), respectively. We notice that second-order correction to the magnon energy in this case becomes positive around the $\Gamma$ point of the BZ in contrast to the previous two processes.\\

\paragraph{\textbf{Fourth process} \underline{$u + u \rightarrow d + u$}\textbf{:}} Now, we focus on the remaining two subdominant processes which include more up magnons in the scattering channel than the down magnons. In this process, there is only one down magnon involved with the matrix element as
\begin{equation}\label{eq.20}
{\cal{V}}^{(4)}_{\vk;\vq,\vp}
=
\left( 
\frac{|\gamma_\vk| + |\gamma_\vq|}{2}
\right) e^{i\zeta}
- 
i|\gamma_{\vp-\vq}| s_\theta
-
i|\gamma_{\vp-\vk}| s_\kappa,
\end{equation}
while the kinematic factor is written down as 
\begin{equation}\label{eq.21}
K^{(4)}_\vk = \int_{\vq\vp} \delta(\epsilon^u_\vk + \epsilon^u_\vq - \epsilon^d_\vp - \epsilon^u_{\vk+\vq-\vp}).
\end{equation}
This process includes three of the scattering magnons in the optical branch. The corresponding decay rate is obtained from Eq.~\eqref{eq.9} as
\begin{equation}\label{eq.22}
W^{(4)}_\vk = \int_{\vq\vp} \frac{J^2|\mathcal{V}^{(4)}_{\vk;\vq,\vp}|^2 {\mathcal F}_{\vk;\vq,\vp}}{16} \delta(\epsilon^u_\vk + \epsilon^u_\vq - \epsilon^d_\vp - \epsilon^u_{\vk+\vq-\vp}).
\end{equation}
Both $K^{(4)}_\vk$, and $W^{(4)}_\vk$, are computed numerically and the corresponding results are shown in panel (b), and (f) of Fig.~\ref{fig:Fig4}, respectively. The energy renormalization due to the real part of the self-energy correction is shown in Fig.~\ref{fig:Fig4}(k) at various temperatures. The self-consistent analysis at 25 K with Hartree correction as before modifies the magntiudes and qualitative features of the decay rate and real part of the self-energy as shown in panel (d) and (i) of Fig.~\ref{fig:Fig5}, respectively. For the optical magnon, temperature factors significantly suppress both the decay rate and the real part. Counterintuitively, this results in the weakest contribution to renormalization corrections, even though it contains an acoustic magnon. \\

\paragraph{\textbf{Fifth process} \underline{$u + u \rightarrow u + u$}\textbf{:}} Despite being the least important process, scattering between magnons in the optical branch surprisingly contributes nearly an order of magnitude more than the previous process ($u + u \rightarrow d + u$) due to subtle variations in the matrix elements, temperature, and kinematic factors, as shown in Fig.~\ref{fig:Fig4}(g,l). The matrix element, in this case, is given by
\begin{equation}\label{eq.23}
{\cal{V}}^{(5)}_{\vk;\vq,\vp}
=
(|\gamma_\vk| + |\gamma_\vq|) c_\zeta
-
|\gamma_{\vp-\vq}| c_\theta
-
|\gamma_{\vp-\vk}| c_\kappa.
\end{equation}
The kinematic factor for this process is dictated by the magnon energies in the optical branch only and is written as 
\begin{equation}\label{eq.24}
K^{(5)}_\vk = \int_{\vq \vp} \delta(\epsilon^u_\vk + \epsilon^u_\vq - \epsilon^u_\vp - \epsilon^u_{\vk+\vq-\vp}).
\end{equation}
The corresponding decay rate is obtained from Eq.~\eqref{eq.9} as 
\begin{equation}\label{eq.25}
W^{(5)}_\vk = \int_{\vq\vp} \frac{J^2|\mathcal{V}^{(5)}_{\vk;\vq,\vp}|^2 {\mathcal F}_{\vk;\vq,\vp}}{16} \delta(\epsilon^u_\vk + \epsilon^u_\vq - \epsilon^u_\vp - \epsilon^u_{\vk+\vq-\vp}).
\end{equation}
The corresponding results for both $K^{(5)}_\vk$ and $W^{(5)}_\vk$ are shown in panels (a) and (g) of Fig.~\ref{fig:Fig4}, respectively. Owing to the gapped structure of the magnon dispersion, this fifth process is expected to be highly suppressed at low temperatures. Nevertheless, it is noteworthy that the relative magnitude of this channel exceeds that of the fourth process, which involves three upper-branch magnons, as evident from the comparison in Fig.~\ref{fig:Fig4}. The real part of the associated self-energy is displayed in Fig.~\ref{fig:Fig4}(l). As in the other scattering channels, the self-consistent treatment yields qualitative changes, as shown in Fig.~\ref{fig:Fig5}(e,j) for the decay rates and real parts, respectively.

We also observe that the kinematic factors for the second, third, and fourth processes, namely, $K^{(2)}_\vk$, $K^{(3)}_\vk$, and $K^{(4)}_\vk$ \--- are identical, while those for the first and fifth processes \--- $K^{(1)}_\vk$ and $K^{(5)}_\vk$ \--- also share the same structure up to an overall scaling factor. Despite these similarities in kinematic structure, the decay rates and the real parts of the self-energy differ across the five channels due to variations in the matrix elements and temperature-dependent occupation factors. 

\begin{figure}[t!]
\centering \includegraphics[width=1.0\linewidth]{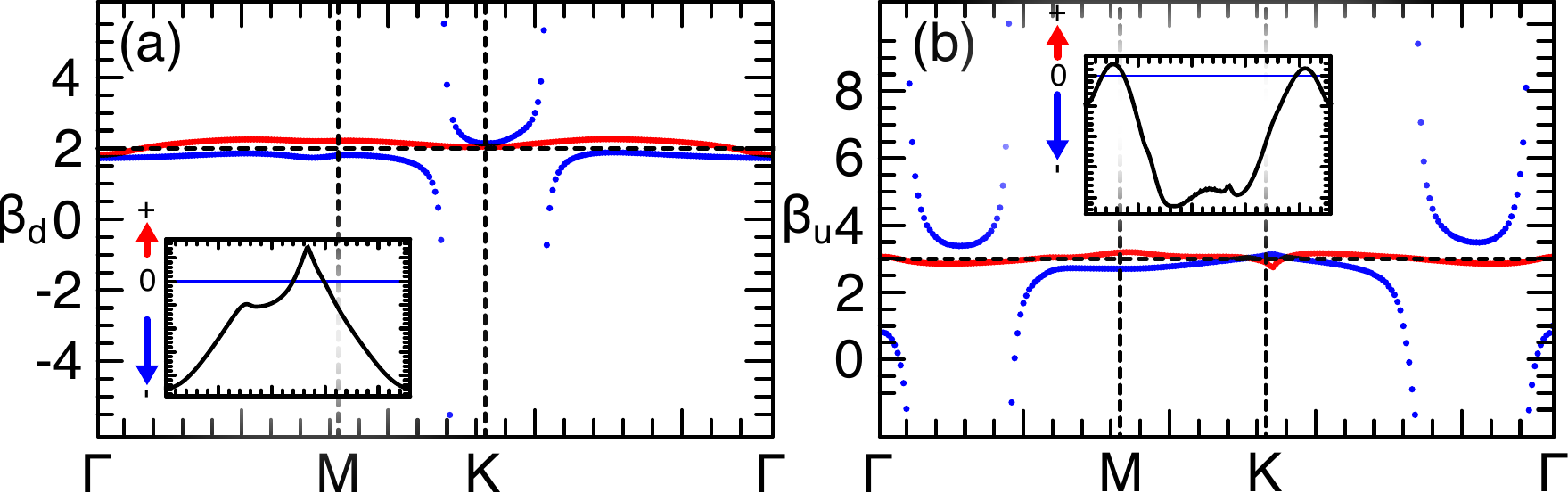}
\caption{The best fit to the temperature power law for the real and the imaginary part of the self-energy. The red (\textit{blue}) curve shows the fitted exponent relevant for the decay rate (\textit{real part}) corresponding to the down [$T^{\beta_d}$, see panel (a))] and the up [$T^{\beta_u}$, see panel (b)] magnon bands.}\label{fig:Fig6}
\end{figure}

\subsection{Temperature dependence \label{sec:Sec_II.III}}

\begin{figure*}[t!]
\centering \includegraphics[width=0.85\linewidth]{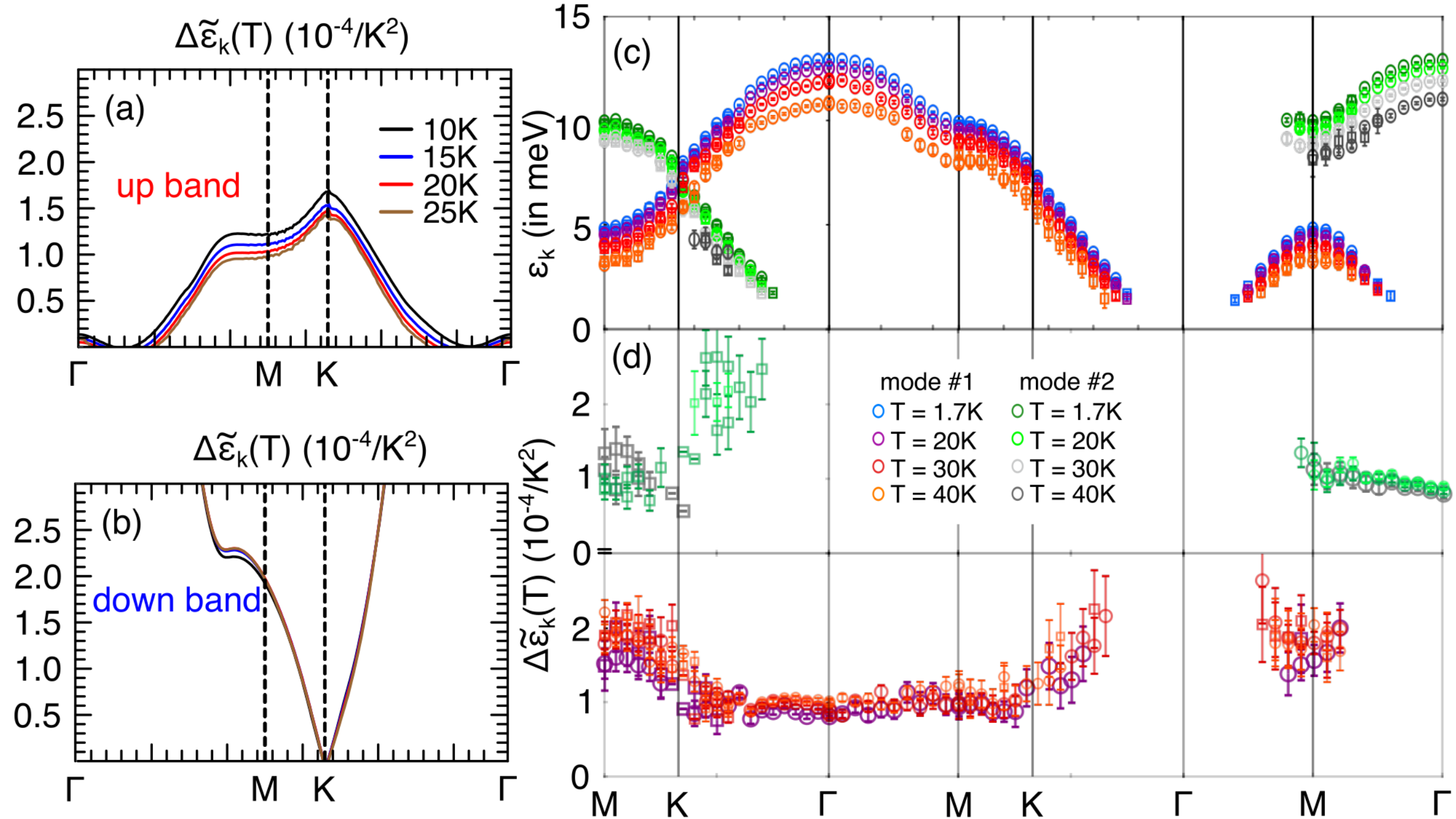}
\caption{(a) The temperature-dependent scaled energy parameter as defined in Eq.~\eqref{Nikitin} for the up band at several temperatures (solid line). (b) The same for the down band. (c) The experimental data set for up and down band, and (d) the corresponding scaled energy $\Delta \tilde{\varepsilon}_\vk(T)$ at various temperatures, reproduced from Ref.~\cite{PhysRevLett.129.127201}.} \label{fig:Fig7}
\end{figure*}

In this section, we examine how the second-order self-energy contributions depends on temperature. We follow a similar approach as outlined in Sec.~\ref{sec.sec_II.I} for the Hartree correction. We then calculate the real and imaginary parts of the self-energy correction for both the up (optical) and down (acoustic) magnon bands at fourteen different temperatures, ranging from 5 K to 31 K. Our objective is to identify a temperature fitting of the form $a T^{\beta}$ for both bands.

For the down band, we find that the best fit for the exponent is $\beta_d = 2$ for both the real and imaginary parts, as illustrated in Fig.~\ref{fig:Fig6}(a). Similarly, for the up band, the best fit is $\beta_u = 3$ for both the real and imaginary parts, shown in panel (b) of Fig.~\ref{fig:Fig6}. While the temperature dependence for the \textit{acoustic} band is consistent with previous theoretical and experimental works, the \textit{cubic} power law for the \textit{optical} band is a new \textit{finding} of our work. We summed over all the contributions from second to fifth process to obtain the self-energy corrections for the optical band. Numerical fitting of the real parts reveals asymptotic characteristics that are numerical artifacts resulting from changes in the sign of the renormalized bands across different regions of the Brillouin zone (see insets of Fig.~\ref{fig:Fig6}).

\section{Contrast with INS experiment \label{sec:sec_III}}

After obtaining the decay rates and self-energy corrections for the five processes in the previous section, we now calculate the total decay rate and self-energy correction for the up and down magnon bands. We notice that the overall magnitudes for the first, second and third processes are much larger than the fourth and fifth processes as shown in Fig.~\ref{fig:Fig4}, and Fig.~\ref{fig:Fig5}. However, for completeness, we consider all relevant processes to facilitate a quantitative comparison with the experimental results reported in Ref.~\cite{PhysRevLett.129.127201}. 

A notable feature highlighted by our exact numerical results in Fig.~\ref{fig:Fig5} is the lack of any van Hove-like singularities in both the renormalized band structure and the decay rates. These singularities were previously predicted based on the thermal magnon approximation used in our earlier theoretical work~\cite{PhysRevX.8.011010}. Consequently, our findings are qualitatively consistent with the recent INS experiment~\cite{PhysRevLett.129.127201} and also other similar experiments. It is important to point out that our model incorporates only nearest-neighbor exchange coupling, whereas the study in Ref.~\cite{PhysRevLett.129.127201} took into account additional neighbor couplings. Nonetheless, we convert our quantitative estimates for the Hartree renormalization and the real part of the self-energy to align with a quantity referred to as ``scaled energy" in the INS study. This scaled energy is defined as follows
\begin{equation}\label{Nikitin}
\Delta \tilde{\varepsilon}_\vk(T)
=
\frac{\varepsilon_\vk(0) - \varepsilon_\vk(T)}{\varepsilon_\vk(0)T^2},
\end{equation}
where $\varepsilon_\vk(T)$ measures the total magnon energy at a finite temperature $T$ by incorporating the Hartree term and the second-order correction to the bare band structure. The factor $T^2$ is introduced in line with the predicted quadratic temperature dependence. In our findings, we obtain $T^2$ dependence for the down magnon band, while the up magnon band appears to exhibit $T^3$ dependence. However, for the sake of comparison with experimental data, we continue to use the same definition of $\Delta \tilde{\varepsilon}_\vk(T)$ for both the bands. Utilizing the data sets for different temperatures for the up and down band, we make an effort to fit the experimental data for $\Delta \tilde{\varepsilon}_\vk(T)$ in Ref.~\cite{PhysRevLett.129.127201} as shown in Fig.~\ref{fig:Fig7}(a,b). Due to the dense set of data points in Ref.~\cite{PhysRevLett.129.127201} [see panel (c), and (d) in Fig.~\ref{fig:Fig7}] with increasing error bar at higher temperature, we do not aim to match our theoretical results in an exact manner. Nevertheless, the overall trend and the magnitudes match surprisingly well with the experimental results \--- especially the observation that the down band suffers more renormalization due to self-energy correction than the up band. A discrepancy exists between our theoretical results and the experimental data. Specifically, the INS data exhibits a clear energy offset ($\sim 1 \times 10^{-4}$/K$^2$), which is not predicted by our perturbation calculations. Furthermore, while our theory and the experimental data both reveal a strong renormalization of the scaled energy for the optical band ($\sim 1.5\times 10^{-4}$/K$^2$), the scaled energy for the acoustic band approaches zero at the Dirac point. This suggests that the renormalization correction for the down band is negligible near the Dirac point, while the up band energy is reduced. We attribute this discrepancy to the omission of vertex corrections in our current calculations. However, the magnitude of this mismatch remains relatively small. Moreover, a quantitative comparison in this case is not appropriate because of the simplified model Hamiltonian adopted in Eq.~\eqref{eq.1}.

Another issue still remains: the emergence of negative renormalization for the acoustic band, as revealed by second-order perturbation theory around the $\Gamma$ point in the BZ [see Fig.~\ref{fig:Fig4}(h) or Fig.~\ref{fig:Fig5}(f)]. Since we are dealing with bosonic excitations, a negative energy correction for the acoustic band near the $\Gamma$ point leads to the presence of negative energy bosons. In contrast, a similar negative correction for the optical band does not present any problems, as it belongs to the optical branch, which maintains a finite positive energy throughout the BZ.

Firstly, the single-ion anisotropy term ($-2AS$) gaps out the down band magnon spectrum at the $\Gamma$ point. Therefore, a negative renormalization remains valid as long as the overall energy stays positive, which holds true at lower temperatures. However, at higher temperatures, the self-energy correction becomes more significant, potentially resulting in a negative overall renormalized spectrum. In fact, the negative renormalization correction can exceed the positive anisotropic energy gap for the down band at elevated temperatures, as indicated in Fig.~\ref{fig:Fig4}(h) and through the self-consistent analysis in Fig.~\ref{fig:Fig5}(f). This occurrence signals a breakdown of the perturbative approximation~\cite{PhysRevB.48.3792,Kosevich1986}. To address this, more sophisticated methods should be employed, such as state-of-the-art quantum Monte Carlo~\cite{PhysRevLett.121.077202,PhysRevB.61.364} or matrix-product state techniques~\cite{PhysRevB.97.235155,Boris2022}. In principle, the negative energy down magnons for the small momentum case (i.e., long-wavelength) correspond to thermal fluctuations, which become increasingly significant as temperature rises in a ferromagnet. This phenomenon is well-known in perturbation theory and has been predicted in previous theoretical work by Kosevich and Chubukov~\cite{Kosevich1986}.

A final point concerns the temperature dependence of the upper band and an apparent inconsistency between our second-order self-energy correction (predicting $T^3$ dependence) and the $T^2$ dependence observed experimentally in Ref.~\cite{PhysRevLett.129.127201}. The experimental fitting in Ref.~\cite{PhysRevLett.129.127201} was performed near the M point in the BZ. We note that the Hartree correction for both bands exhibits approximately quadratic temperature dependence and, near the M point, its magnitude exceeds that of the second-order contributions. Therefore, the quadratic behavior is expected to mask the cubic corrections in the next leading order, making the observation of the subleading cubic dependence experimentally challenging. \\

\section{Comparison to Bravais case: Triangular ferromagnet \label{sec:sec_IV}}

To clarify the influence of the hexagonal BZ on magnon-magnon scattering and analyze the limitations of the ``thermal magnon approximation" at low temperatures, we consider a simplified one-band model on a triangular lattice with nearest-neighbor interactions. We begin by analyzing the renormalization using the full four-dimensional self-energy integration, and conclude with a comparison between the Bravais and non-Bravais scenarios.

We start with a ferromagnetic Heisenberg model as in Eq.~\eqref{eq.1} with a single-ion anisotropy relevant for the triangular lattice. From a materials standpoint, the isotropic Heisenberg ferromagnet on a triangular lattice serves as a reasonable model for monolayer MnBi$_2$Te$_4$, where long-range magnetic order in two dimensions is stabilized by a single-ion anisotropy (SIA) arising from spin–orbit coupling on Mn and Te atoms~\cite{YangLi_PRB2019_SIA_MnBi2Te4}. In contrast to CrBr$_3$, Mn$^{2+}$ carries a larger spin, $S = \tfrac{5}{2}$, making perturbative treatments even more relevant. For monolayer MnBi$_2$Te$_4$, in Ref.~\cite{YiqunLiu_PRB2023_MnBi2Te4} the ab initio exchange parameters were found  as $J_1 =0.2496$ meV, $J_2 = -0.024$ meV and $J_3 = -0.010$ meV, corresponding to first-, second-, and third-neighbor couplings, respectively. The SIA, arising from spin-orbit interactions on Mn and Te, is given by $E_g = -2AS = 0.144$ meV~\cite{YangLi_PRB2019_SIA_MnBi2Te4}. For the subsequent analysis, we only retain $J_1 \equiv J$ and ignore all the other weak anti-ferromagnetic longer range exchange couplings. The intra-layer Curie temperature is reported to be around $15.2$ K~\cite{PhysRevX.11.011003}.

\begin{figure}[t!]
\centering \includegraphics[width=1.0\linewidth]{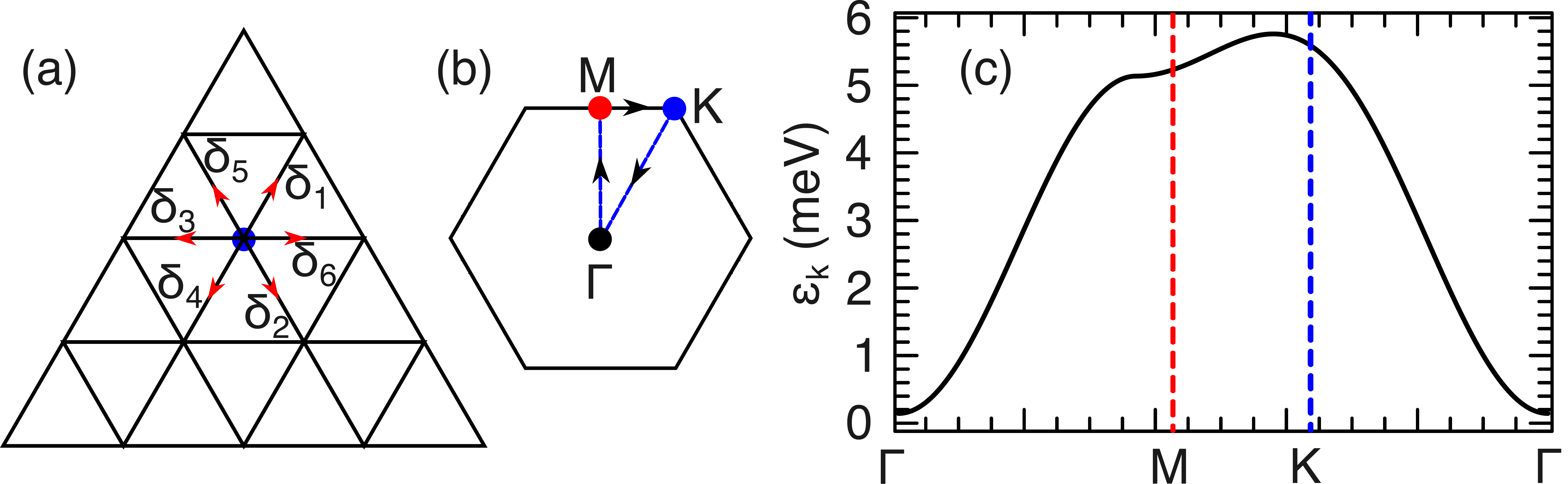}
\caption{(a) Schematic of the monopartite triangular lattice with its six nearest-neighbor vectors. (b) The corresponding Brillouin zone (BZ) showing the high-symmetry points. (c) The magnon spectrum obtained using linear spin-wave theory along the high-symmetry path. The coordinates of the key high-symmetry points are: K = ($\tfrac{2\pi}{3a},\tfrac{2\pi}{\sqrt{3}a}$), and M = ($0,\tfrac{2\pi}{\sqrt{3}a}$), where $a$ denotes the lattice constant.}\label{fig:Fig8}
\end{figure}

Applying the Holstein-Primakoff transformation to the ferromagnetic Heisenberg model (see Eq.~\eqref{eq.1}) on a triangular lattice, we first obtain the effective linear spin wave part of the Hamiltonian as
\begin{equation}\label{eq.26}
\mathcal{H}_0 = \sum_{\vk} \varepsilon_\vk  a^\dag_\vk a_\vk,
\end{equation}
where $a^\dag_\vk$ creates a spin wave excitation (magnon) with momentum $\vk$ and spectrum as\
\begin{equation}\label{eq.27}
\varepsilon_\vk = JS(6 - \eta_\vk) - 2AS,
\end{equation}
where $\eta_\vk = \sum_{i=1}^{6} e^{i \vk \cdot {\bm{\delta}}_i}$ with $\bm{\delta}_i$ denoting the six nearest neighbour vectors on the triangular lattice, and $A$ is the SIA. Here, we choose $\bm{\delta}_1$, $\bm{\delta}_2$, and $\bm{\delta}_3$ as before, and other vectors are chosen as $\bm{\delta}_4 = -\bm{\delta}_1$, $\bm{\delta}_5 = -\bm{\delta}_2$, and $\bm{\delta}_6 = - \bm{\delta}_3$ as illustrated in Fig.~\ref{fig:Fig8}(a). The corresponding spectrum is shown along the high symmetry points in Fig.~\ref{fig:Fig8}(c). Note that there is only one species of magnons for the monopartite triangular lattice. Now, going beyond the linear spin wave theory, we can obtain the quartic interacting Hamiltonian as
\begin{align}
\nonumber
\mathcal{H}_{1} 
=
\frac{J}{8N} 
\sum_{\vk, \vq, \vp} 
\big[ 2 \eta_\vk \, + \, & \eta_{\vk+\vq-\vp} \, + \, \\ 
\label{eq.28}
&
\, \eta_\vp - 4 \eta_{\vq - \vp}\big] 
a^{\dagger}_\vp a^{\dagger}_{\vk+\vq-\vp} a_{\vq} a_{\vk}.
\end{align}
Note that, we labeled the various momentum indices in the same manner as in Eq.~\eqref{eq.4}, however, we chose the high-symmetry path along a different sector as shown in Fig.~\ref{fig:Fig8}(b). Here, again, the single-ion anisotropy has been neglected in the derivation of the interaction vertex. 

\subsection{Hartree contribution in triangular ferromagnet\label{sec:sec_IV.I}}

\begin{figure}[t!]
\centering \includegraphics[width=1.0\linewidth]{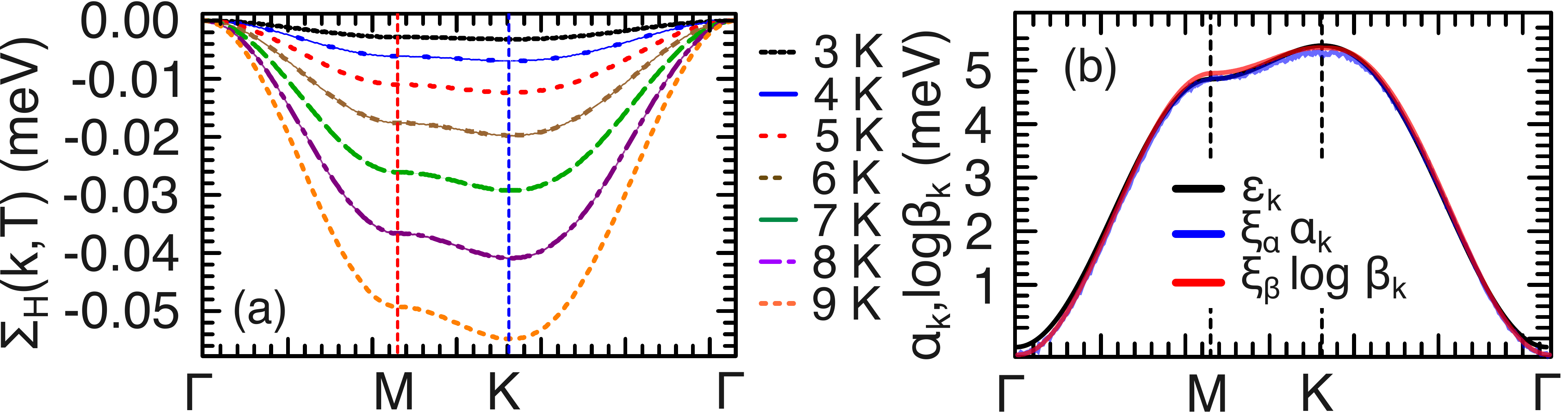}
\caption{(a) The evolution of the Hartree correction (see Eq.~\eqref{eq.29}) to the magnon energy renormalization in case of a the triangular lattice ferromagnet at different temperatures. Notice that the overall contribution is negative and grows as a function of temperature. The bare (solid line) and self-consistent (dashed) results coincide. (b) The best fitting of the parameters $\alpha_\vk$, and $\beta_\vk$ in for the Hartree contribution as assumed as $\Sigma_{\rm H}(\vk,T) = \alpha_\vk T^2|\ln (\beta_\vk T)|$. Here $\alpha_\vk (\ln \beta_\vk)$ are proportional to the bare dispersion by pre-factors $\xi_\alpha (\xi_\beta)$, \textit{i.e.}, $\xi_a \alpha_\vk \approx \varepsilon_\vk$, and $\xi_b \ln \beta_\vk \approx \varepsilon_\vk$ with $\varepsilon_\vk$ being the bare dispersion in Eq.~\eqref{eq.27}. Here, $\xi_{a,b}$ are some constants in arbitrary unit.}\label{fig:Fig9}
\end{figure}

At first, we consider the Hartree approximation on the interaction vertex in Eq.~\eqref{eq.28}. This corresponds to a similar Feynmann diagram as in Fig.~\ref{fig:Fig2}(a) for the honeycomb case. Consequently, the Hartree self-energy reads as
\begin{equation}\label{eq.29}
\Sigma_{\rm H}(T) 
= 
\sum_\vk \tilde{h}_\vk(T) a^\dag_\vk a_\vk,
\end{equation}
where $\tilde{h}_\vk(T)$ is defined as
\begin{equation}\label{eq.30}
\tilde{h}_\vk(T) = J \sum_\vq f(\varepsilon_\vq) \left(\eta_\vk + \eta_\vq - \eta_{\vk-\vq} - 6 \right),
\end{equation}
with $f(x)$ denoting the Bose-Einstein distribution function as before. The Hartree contribution in this case is structurally similar to that in Eq.~\eqref{eq.6}. However, after numerically evaluating the integral in Eq.~\eqref{eq.30}, we find that the result is well-approximated by $\Sigma_{\rm H}(\vk, T) \approx -\alpha_\vk T^2 |\ln (\beta_\vk T)|$, where $\alpha_\vk (\ln \beta_\vk)$ is proportional to the non-interacting dispersion with some proportionality constant $\xi_\alpha (\xi_\beta)$ (which are independent of temperatures and momentum and depend only on the material parameters such as $J$, $S$, and $A$) as shown in Fig.~\ref{fig:Fig9}(b). This temperature dependence is very distinct from the behavior found for the honeycomb lattice, suggesting a generic distinction between Bravais and non-Bravais lattice systems. Moreover, we perform both bare and self-consistent analysis of $\Sigma_{\rm H}(T)$. Apart from overall magnitude, the qualitative features in both the cases, remain identical as is shown in Fig.~\ref{fig:Fig9}. 

\subsection{Second order correction in triangular ferromagnet\label{sec:sec_IV.II}}

In this section, we perform the second-order self energy correction in the similar spirit to Sec.~\ref{sec.sec_II.II}. A recent theoretical work~\cite{YiqunLiu_PRB2023_MnBi2Te4} carried out such a perturbation theory in the framework of the thermal magnon approximation adopted in our previous work~\cite{PhysRevX.8.011010}, and obtained respective magnon renormalization with apparent van Hove like features around K and M points in the BZ. Here we carry out the full four-dimensional integration without adhering to any thermal approximation, similar to the honeycomb case. The corresponding self-energy, similar to one in Eq.~\eqref{eq.9}, reads as follows
\begin{equation}\label{eq.31}
\Sigma_{\rm S}(\omega,\vk)
=
\frac{J^2}{64} \int_{\vq\vp}
\frac{{|\cal{V}_{\vk;\vq,\vp}|}^2 \mathcal{F}_{\vk;\vq,\vp}}{\omega + \varepsilon_\vq - \varepsilon_\vp - \varepsilon_{\vk+\vq-\vp} + i\delta},
\end{equation}
where the temperature factor $\mathcal{F}_{\vk;\vq,\vp}$ is defined as before in Eq.~\eqref{beq.7}, and the matrix element $\mathcal{V}_{\vk;\vq,\vp}$ is defined as in Eq.~\eqref{eq.28}. The second-order correction $\Sigma_{\rm{S}}(\varepsilon_\vk,\vk)$ at the K and M point indeed qualitatively differs, without any presence of van Hove like features around K and M point in the BZ, in comparison to Ref.~\cite{YiqunLiu_PRB2023_MnBi2Te4}. 

\begin{figure}[t!]
\centering \includegraphics[width=1.0\linewidth]{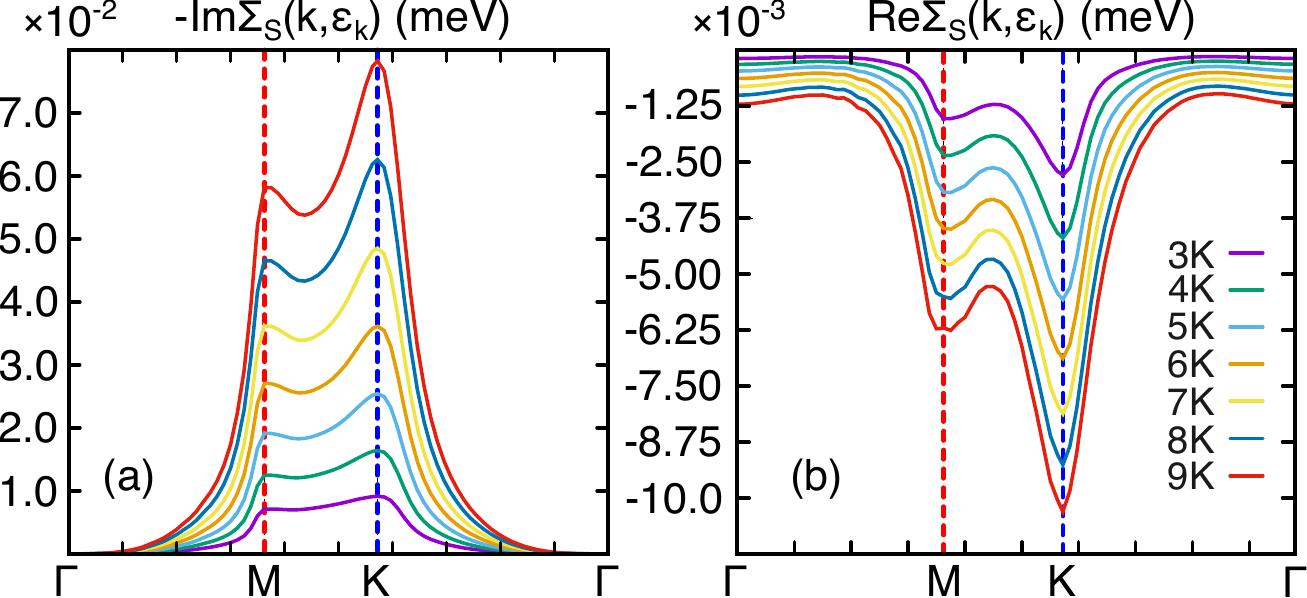}
\caption{The imaginary (a) and real (b) part of the second-order self-energy correction obtained from Eq.~\eqref{eq.31} at various temperatures along the high-symmetry points in the BZ. Note, in particular, the apparent absence of any van Hove singularities around K and M point in the BZ. We chose the parameters as $J = 0.2496$ meV and $A = -0.028$ meV.}\label{fig:Fig10}
\end{figure}

\begin{figure}[t!]
\centering \includegraphics[width=1.0\linewidth]{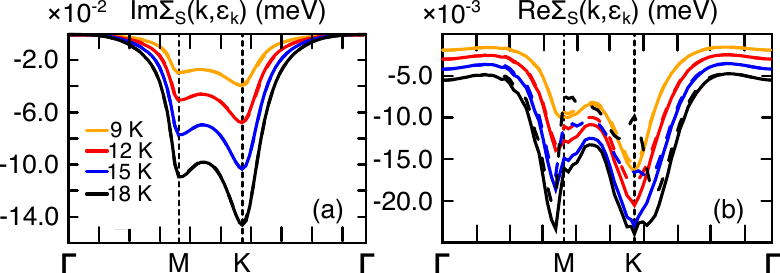}
\caption{A comparison between the bare (solid line) and self-consistent (dashed line) calculation for the imaginary (a) and real (b) part of the second-order self-energy correction obtained from Eq.~\eqref{eq.31} at various temperatures along the high-symmetry points in the BZ. We chose the parameters as $J = 0.2496$ meV and $A = -0.028$ meV.}\label{fig:Fig11}
\end{figure}

The real and imaginary parts of the self-energy correction for different temperatures, are shown in panels (b), and (a) of Fig.~\ref{fig:Fig10}, respectively. We further note that the real part of the self-energy is negative and grows in magnitude away from the $\Gamma$ point in the BZ. This is consistent with what we obtained in the case of a honeycomb lattice. In this case also, the negative contribution is valid as long as the total magnitude of the renormalized magnon energy remains positive in the presence of the SIA. In contrast to the honeycomb case, the self-consistent second-order self-energy correction does not lead to much qualitative modifications in the profile compared to bare self-energy, as shown in Fig.~\ref{fig:Fig11}(a,b). For completeness, we go beyond the Curie temperature ($T > 15$ K) where there is a significant modification in the real part of the self-consistent self-energy. This is in contrast to the decay rate for which self-consistent and non-self-consistent calculations are indistinguishable on the scale of the plot.

\section{Discussion and conclusion \label{sec:sec_V}}

In this paper, we investigated the role of magnon-magnon interactions in the ferromagnetic material CrBr$_3$. Our focus was on the temperature-dependent renormalization of the magnon spectrum and decay rates, using second-order perturbation theory. Our numerical results clearly show that, for the honeycomb lattice, a detailed analysis \--- without any thermal magnon approximations \--- reveals a $T^2$ dependence for the band corresponding to the Hartree contribution. In contrast, the second-order self-energy correction yields a $T^2$ dependence for the lower energy band while the upper energy band exhibits a distinct $T^3$ dependence. These findings starkly contrast with previous predictions, which suggested a quadratic temperature dependence for both bands in the cases of Hartree and second-order corrections.

In this context, we emphasize a subtle point first theoretically predicted by Y. A. Kosevich and A. V. Chubukov in Ref.~\cite{Kosevich1986}. Their analysis evaluated both Hartree and second-order self-energy corrections using a long-wavelength approximation for low-energy magnons in Bravais lattices. This approach predicted additional logarithmic factors: specifically, the Hartree correction behaves as $\Delta \varepsilon^{\rm H}_{\vk} \sim T^2 |\ln \{ {\rm max} (\tfrac{k_{\rm B}T}{JS}, \vk^2)\}|$, while the second-order correction scales as $\Delta \varepsilon^{(2)}_{\vk} \sim T^2 |\ln (\tfrac{\vk^2 JS}{k_{\rm B} T})|$ for the regime where $JS \vk^2 \ll k_{\rm B} T$. Our numerical results reveal a similar logarithmic dependence for the triangular lattice case at the Hartree level of perturbation theory, while the second-order correction shows similar characteristics to those previously mentioned in Ref.~\cite{Kosevich1986}. 

Notably, this temperature dependence is absent in non-Bravais lattices. To our knowledge, this work is the first to clearly differentiate between Bravais and non-Bravais lattices in this context. We propose that the presence of Dirac nodes in the band structure underlies the stark difference in temperature behavior between honeycomb (non-Bravais) and triangular (Bravais) magnets. As a secondary observation, we note a slight mismatch between the renormalized optical and acoustic magnon bands at the Dirac point for the honeycomb lattice, which is not reflected in the corresponding decay spectrum. We attribute this discrepancy to the omission of vertex corrections in our current perturbative treatment; their self-consistent inclusion is expected to restore the Dirac cone structure. Nevertheless, we anticipate that the qualitative features of the real part of the self-energy (Figs.~\ref{fig:Fig4} and \ref{fig:Fig5}) will remain robust.

Our work demonstrates the necessity of explicitly calculating all quantities in second-order perturbation theory, without relying on simplifying assumptions such as the thermal magnon or long-wavelength approximations. Applying this methodology to the honeycomb lattice ferromagnet CrBr$_3$ and triangular lattice ferromagnet MnBi$_2$Te$_4$ revealed significant qualitative differences compared to calculations employing the thermal magnon approximation. A detailed comparison with recent INS data from S. E. Nikitin et al.~\cite{PhysRevLett.129.127201} (presented in Sec.~\ref{sec:sec_III}) supports the validity of our approach, as evidenced by the absence of van Hove singularities in our full calculation (Fig.~\ref{fig:Fig4}) \--- consistent with experimental observations for CrBr$_3$. We find a cubic ($T^3$) temperature dependence for the renormalized optical magnon band in the honeycomb lattice, contradicting with the quadratic behavior reported in Ref.~\cite{PhysRevLett.129.127201}. However, it is crucial to note that the total renormalization of the upper band arises from both Hartree and second-order self-energy contributions. The Hartree term, exhibiting a quadratic ($T^2$) dependence (as discussed in Sec.~\ref{sec.sec_II.II}), dominates the second-order self-energy term, thus explaining the observed $T^2$ behavior in experiments. We note that a previous theoretical work~\cite{PhysRevB.111.094430} reports similar $T^3$ correction for both magnon branches, while we find cubic (quadratic) dependence for the optical (acoustic) branch only.

\section{Acknowledgments \label{sec:sec_VI}}

We thank Stanislav E. Nikitin for helping with data interpretation in Ref.~\cite{PhysRevLett.129.127201}, and providing the raw data for Fig.~\ref{fig:Fig7}(c,d). We gratefully acknowledge support from the Office of Basic Energy Sciences, Material Sciences and Engineering Division, U.S. DOE, under Contract No. DE-FG02-99ER45790. Computations were performed on the Amarel cluster at Rutgers University, USA, and University Computer Centre of the University of Greifswald, Germany. \\

\section{Author Contributions \label{sec:sec_VII}}

S.B. developed the idea, S.B. and S.H. performed the analytical and numerical calculations and wrote the manuscript.

\appendix

\section{Hartree approximation \label{sec:sec_app.I}}

Here, we provide the derivation details in Eq.~\eqref{eq.6}. Utilizing the Holstein-Primakoff transformation as defined in Eq.~\eqref{eq.2}, we obtain the quartic part of bosonic Hamiltonian from Eq.~\eqref{eq.1} as
\begin{widetext}
\begin{equation}\label{aeq.1}
\mathcal{H_1} = \frac{J}{4N} \sum_{\{\vk_i\}}
\left(
\gamma^\ast_{\vk_2} a^\dag_{\vk_1} b^\dag_{\vk_2} a_{\vk_3} a_{\vk_4}
+
\gamma_{\vk_4} a^\dag_{\vk_1} a^\dag_{\vk_2} a_{\vk_3} b_{\vk_4}
+
\gamma_{\vk_2} b^\dag_{\vk_1} a^\dag_{\vk_2} b_{\vk_3} b_{\vk_4} 
+
\gamma^\ast_{\vk_4} b^\dag_{\vk_1} b^\dag_{\vk_2} b_{\vk_3} a_{\vk_4}
-
4\gamma_{\vk_4-\vk_2} a^\dag_{\vk_1} b^\dag_{\vk_2} a_{\vk_3} b_{\vk_4}
\right),
\end{equation}
\end{widetext}
where we expanded the boson operators on a sublattice basis and ignored the single-ion anisotropy term. Considering the Hartree diagram in Fig.~\ref{fig:Fig2}(a), we perform bilinear Wick contractions for each term in Eq.~\eqref{aeq.1}. The corresponding expectation values are computed with the ferromagnetic ground state. For example, the first term is expanded as follows
\begin{widetext}
\begin{equation}\label{aeq.2}
\gamma^\ast_{\vk_2} a^\dag_{\vk_1} b^\dag_{\vk_2} a_{\vk_3} a_{\vk_4}
\rightarrow
\gamma^\ast_{\vk_2}
\left[
a^\dag_{\vk_1} \braket{b^\dag_{\vk_2} a_{\vk_3}} a_{\vk_4}
+
a^\dag_{\vk_1} \braket{b^\dag_{\vk_2} a_{\vk_4}} a_{\vk_3}
+
b^\dag_{\vk_2} \braket{a^\dag_{\vk_1} a_{\vk_3}} a_{\vk_4}
+
b^\dag_{\vk_2} \braket{a^\dag_{\vk_1} a_{\vk_4}} a_{\vk_3}
\right].
\end{equation}
\end{widetext}
Now, we rewrite the bilinear expectation values in terms of the diagonal operators as
\begin{subequations}
\begin{align}
\label{aeq.3.1}
\braket{a^\dag_\vk a_\vk} & 
= \frac{\braket{u^\dag_\vk u_\vk} + \braket{d^\dag_\vk d_\vk} + \braket{u^\dag_\vk d_\vk} + \braket{d^\dag_\vk u_\vk}}{2}, \\
\label{aeq.3.2}
\braket{b^\dag_\vk b_\vk} & 
= \frac{\braket{u^\dag_\vk u_\vk} + \braket{d^\dag_\vk d_\vk} - \braket{u^\dag_\vk d_\vk} - \braket{d^\dag_\vk u_\vk}}{2}, \\
\label{aeq.3.3}
\braket{b^\dag_\vk a_\vk} & 
= e^{i\phi_\vk}\frac{\braket{d^\dag_\vk d_\vk} + \braket{d^\dag_\vk u_\vk} -\braket{u^\dag_\vk u_\vk} - \braket{u^\dag_\vk d_\vk}}{2}.
\end{align}
\end{subequations}
Subsequently, in thermal equilibrium, we replace the bilinears as $\braket{u^\dag_\vk u_\vk} = f(\epsilon^u_\vk)$, and $\braket{d^\dag_\vk d_\vk} = f(\epsilon^d_\vk)$ for the up and the down bands respectively. Since the ferromagnetic ground state preserves the total number of magnons, we ignore $\braket{u^\dag_\vk d_\vk}$. Performing the similar Wick contraction for all the five terms in Eq.~\eqref{aeq.1}, and following a simple algebra, we obtain the expression for $f(T)$, and $g_\vk(T)$ in Eq.~\eqref{eq.7}.

\section{Second-order approximation \label{sec:sec_app.II}}

In this section, we present a concise derivation of the second-order perturbation theory leading to the self-energy correction in Eq.~\eqref{eq.9}. While the derivation closely follows our previous work in Ref.~\cite{PhysRevX.8.011010}, we outline the key algebraic steps here for completeness. First, we write down the bare Green's functions for the up and down magnons as
\begin{subequations}
\begin{align}
\label{beq.1.1}
G^u_0(i\omega_n,\vk) 
	& = \frac{1}{i\omega_n - \epsilon^u_{\vk}}, 
		\quad 
		\begin{tikzpicture}[baseline={([yshift=-2pt]current bounding box.center)}]
        	\draw[ultra thick, dashed, -{latex}] (0,0) -- (1,0); 
        \end{tikzpicture}  \\
\label{beq.1.2}
G^d_0(i\omega_n,\vk)
	& = \frac{1}{i\omega_n - \epsilon^d_{\vk}},
		\quad  
		\begin{tikzpicture}[baseline={([yshift=-2pt]current bounding box.center)}]
        	\draw[ultra thick, -{latex}] (0,0) -- (1,0); 
    	\end{tikzpicture}
\end{align}
\end{subequations}
where $\omega_n = \tfrac{2n\pi}{\beta}$ with $\beta = \tfrac{1}{k_{\rm{B}}T}$ is the inverse temperature. Next, note that by rewriting the four magnon interaction terms from the sub-lattice basis [see Eq.~\eqref{aeq.1}] in the diagonal basis, we obtain five distinct scattering channels as illustrated in Eq.~\eqref{eq.4}. Using Feynman diagrams relevant for the third process in Eq.~\eqref{eq.4}, we now illustrated the self-energy corrections from the sunset diagram in Fig.~\ref{fig:Fig2}(b). The latter is obtained as
\begin{align}
&
\nonumber
\Sigma^{(3)}_{\rm{S}}(i\Omega, \vk)  \\
&
\nonumber
=
\int \frac{d\vq d\vp}{\beta^2}
\sum_{\{ i\omega_{i} \}} 
\underbrace{\frac{|{\mathcal{V}}^{(3)}_{\vk;\vq,\vp}|^2}{
(i\omega_1 - \epsilon^d_\vp)
(i\omega_2 - \epsilon^d_{\vk + \vq - \vp})
(i\omega_3 - \epsilon^d_\vq)}}_{\textbf{C}} \\
&
\label{beq.2}
\Rightarrow
\underbrace{
\frac{|{\mathcal{V}}^{(3)}_{\vk;\vq,\vp}|^2}
{(i\omega_1 - \epsilon^d_\vp)
(i\omega_2 - \epsilon^d_{\vk+\vq-\vp})
(i\omega_1 + i\omega_2 - i\Omega - \epsilon^d_\vq)
}
}_\textbf{C},
\end{align}
where $\Omega + \omega_3= \omega_1 + \omega_2$ is evident from energy conservation at each vertex. The Matsubara summation can be performed easily as follows
\begin{widetext}
\begin{align}\label{beq.3}
\nonumber
&
\sum_{\{i \omega_i\}}\textbf{C}
= 
\sum_{\{ i\omega_i \}} 
\frac{1}{i\omega_1 - \epsilon^d_\vp}
\frac{1}{i\omega_2 - \epsilon^d_{\vk+\vq-\vp}}
\frac{1}{i(\omega_1 + \omega_2 - \Omega) - \epsilon^d_\vq} \\
\nonumber
&
=
\sum_{\{ i\omega_i \}} 
\frac{1}{i\omega_2 - \epsilon^d_{\vk+\vq-\vp}}
\Bigl[ 
\frac{1}{i\omega_1 - \epsilon^d_\vp}
-
\frac{1}{i(\omega_1 + \omega_2 - \Omega) - \epsilon^d_\vq}
\Bigr] 
\Bigl[
\frac{1}{i(\omega_2 - \Omega)-(\epsilon^d_\vq  - \epsilon^d_\vp)}
\Bigr] \\
&
=
\underbrace{
\sum_{\{ i\omega_i \}} 
\frac{1}{i\omega_2 - \epsilon^d_{\vk+\vq-\vp}} 
\frac{1}{i(\omega_2-\Omega)-(\epsilon^d_\vq - \epsilon^d_\vp)}
\frac{1}{i\omega_1 - \epsilon^d_\vp}
}_{\mathbf{A}}
-
\underbrace{
\frac{1}{(i\omega_2 - \epsilon^d_{\vk+\vq-\vp})} 
\frac{1}{i(\omega_2-\Omega) - (\epsilon^d_\vq - \epsilon^d_\vp)} 
\frac{1}{i(\omega_1 + \omega_2 - \Omega) - \epsilon^d_\vq}
}_{\mathbf{B}}.
\end{align}
\end{widetext}
Using the standard formula for Matsubara summation, we obtain
\begin{subequations}
\begin{align}
&
\label{beq.4.1}
{\mathbf{A}}
=
\beta^2 
\frac{f(\varepsilon_\vp) f(\epsilon^d_\vq - \epsilon^d_\vp)
-
f(\varepsilon_\vp) f(\epsilon^d_{\vk+\vq-\vp})}
{i\Omega - (\epsilon^d_\vp + \epsilon^d_{\vk+\vq-\vp} - \epsilon^d_\vq)}, \\
&
\label{beq.4.2}
{\mathbf{B}}
=
\beta^2 
\frac{f(\epsilon^d_\vq) f(\epsilon^d_\vq - \epsilon^d_\vp)
-
f(\epsilon^d_{\vk+\vq-\vp}) f(\epsilon^d_\vq)}
{i\Omega - (\epsilon^d_\vp + \epsilon^d_{\vk+\vq-\vp} - \epsilon^d_\vq)},
\end{align}
\end{subequations}
where $f(x)$ is the Bose-Einstein distribution function. Adding $\mathbf{A}$, and $\mathbf{B}$ in Eq.~\eqref{beq.3}, we obtain the self-energy correction as
\begin{widetext}
\begin{equation} \label{beq.5}
\Sigma^{(3)}_{\rm{S}}(i\Omega,\vk)
=
\int d\vq d\vp 
|{\cal{V}}^{(3)}_{\vk;\vq,\vp}|^2 
\frac{
\Bigl[
f(\epsilon^d_\vq - \epsilon^d_\vp) 
-
f(\epsilon^d_{\vk+\vq-\vp})
\Bigr]
\Bigl[
f(\epsilon^d_\vp) - f(\epsilon^d_\vq)
\Bigr]}
{i\Omega - (\epsilon^d_\vp + \epsilon^d_{\vk+\vq-\vp} - \epsilon^d_\vq)}.
\end{equation}
\end{widetext}
We now perform an analytic continuation of $\{ i\Omega \rightarrow \Omega + i\delta \}$ where $\delta$ is a small positive number. Consequently, the self-energy correction is simplified as 
\begin{widetext}
\begin{equation}\label{beq.6}
\Sigma^{(3)}_{\rm{S}}(\Omega, \vk)
=
\int d\vq d\vp
|{\cal{V}}^{(3)}_{\vk;\vq,\vp}|^2 
\frac{
\Bigl[
f(\epsilon^d_\vq - \epsilon^d_\vp) 
-
f(\epsilon^d_{\vk+\vq-\vp})
\Bigr]
\Bigl[
f(\epsilon^d_\vp) - f(\epsilon^d_\vq)
\Bigr]}
{\Omega - (\epsilon^d_\vp + \epsilon^d_{\vk+\vq-\vp} - \epsilon^d_\vq) + i\delta}
=
\int d\vq d\vp
\frac{|{\cal{V}}^{(3)}_{\vk;\vq,\vp}|^2 \mathcal{F}_{\vk;\vq,\vp}}{\Omega + \epsilon^d_\vq - \epsilon^d_\vp - \epsilon^d_{\vk+\vq-\vp} + i\delta}.
\end{equation}
\end{widetext}
Simplifying further on the Bose-Einstein factors, we obtain the exact form as written in Eq.~\eqref{eq.10}. For the other four processes, as written in Eq.~\eqref{eq.4}, the sunset diagram is modified corresponding to the underlying magnon branches, and the corresponding dispersions enter into Eq.~\eqref{eq.9}. The temperature factor $\mathcal{F}_{\vk;\vq,\vp}$ can be further simplified as 
\begin{widetext}
\begin{equation}\label{beq.7}
\mathcal{F}_{\vk;\vq,\vp}
=
f(\varepsilon_\vq) + f(\varepsilon_\vq) f(\varepsilon_\vp) + f(\varepsilon_\vq) f(\varepsilon_{\vk+\vq-\vp}) - f(\varepsilon_\vp) f(\varepsilon_{\vk+\vq-\vp}).
\end{equation}
\end{widetext}

\bibliography{References}

\end{document}